\documentclass{emulateapj}
\usepackage{apjfonts}
\usepackage{natbib}

\shorttitle{AGN IN CLUSTERS OF GALAXIES} 
\shortauthors{MARTINI, MULCHAEY, \& KELSON} 
\slugcomment{ApJ accepted [18 April 2007]}

\newcommand{\eg}{{\rm e.g.}}

\newcommand{\ergs}{erg s$^{-1}$}
\newcommand{\kms}{km s$^{-1}$}
\newcommand{\chandra}{{\it Chandra}}

\begin{document}

\title{The Distribution of Active Galactic Nuclei in Clusters of Galaxies}  

\author{Paul Martini} 

\affil{Department of Astronomy, The Ohio State University, 
140 West 18th Avenue, Columbus, OH 43210, martini@astronomy.ohio-state.edu}

\author{John S. Mulchaey, Daniel D. Kelson} 

\affil{Carnegie Observatories, 813 Santa Barbara St., Pasadena, CA 91101-1292} 

\begin{abstract}

We present a study of the distribution of AGN in clusters of galaxies with a 
uniformly selected, spectroscopically complete sample of 
35 AGN in eight clusters of galaxies at $z = 0.06 \rightarrow 0.31$. 
We find that the 12 AGN with $L_X > 10^{42}$ \ergs\ in cluster members 
more luminous than a rest-frame $M_R < -20$ mag are more centrally 
concentrated than typical cluster galaxies of this luminosity, although these 
AGN have comparable velocity and substructure distributions to other cluster 
members. 
In contrast, a larger sample of 30 cluster AGN with $L_X > 10^{41}$ \ergs\ do 
not show evidence for greater central concentration than inactive cluster 
members, nor evidence for a different kinematic or substructure distribution. 
As we do see clear differences in the spatial and kinematic distributions of 
the blue Butcher-Oemler and red cluster galaxy populations, any 
difference in the AGN and inactive galaxy population must be less distinct than 
that between these two pairs of populations. 
Comparison of the AGN fraction selected via X-ray emission in this study to 
similarly-selected AGN in the field indicates that the AGN fraction is not 
significantly lower in clusters, contrary to AGN identified via 
visible-wavelength emission lines, but similar to the approximately constant 
radio-selected AGN fraction in clusters and the field. We also find 
significant evidence for variation in the AGN fraction between 
clusters and explore the dependence of cluster AGN fraction on redshift, 
velocity dispersion, amount of cluster substructure, and fraction of 
Butcher-Oemler galaxies. While we see weak evidence for several trends, 
there are several correlations between these four parameters in our small 
sample of eight clusters that preclude identification of which one(s) most 
strongly influence the cluster AGN fraction. 

\end{abstract}


\section{Introduction}

Early work on clusters of galaxies found that emission-line galaxies and 
active galactic nuclei (AGN) were rarer in cluster members than in regions 
of lower galaxy density \citep{osterbrock60,gisler78,dressler85}. 
Extensive work on cluster galaxy populations 
demonstrated that cluster galaxies were dominated by early-type galaxies 
with old stellar populations and the fraction of these quiescent galaxies 
increased with galaxy surface density \citep{dressler80}. These differences 
are now ascribed to multiple physical mechanisms that cause cluster members to 
possess less of the cold gas necessary for young star formation and accretion 
onto supermassive black holes.

The relative rarity of AGN in clusters of galaxies made it difficult 
to acquire sufficiently large samples for demographic studies. 
For example, \citet{dressler85} obtained spectra of over 1000 cluster members 
in 10 low-redshift clusters of galaxies and identified AGN in only 1\% of 
them. This low fraction discouraged large spectroscopic studies whose primary 
purpose was to identify AGN in clusters of galaxies. However, the advent of 
the {\it Chandra X-Ray Observatory} has provided a new means to efficiently 
identify AGN in clusters. Motivated by early evidence of an excess of 
X-ray point sources in the fields of rich clusters of galaxies 
\citep{cappi01,sun02,molnar02}, we \citep{martini02} obtained spectra of the 
bright counterparts to 
X-ray sources in the field of the rich cluster Abell~2104 ($z=0.154$). 
This study identified six bright cluster galaxies coincident with 
luminous X-ray sources ($L_X > 10^{41}$ \ergs), or approximately 5\% of 
cluster members more luminous than $M_R = -20$ mag. While only the most 
X-ray luminous of these six galaxies have the emission-line diagnostics of 
AGN, the remaining X-ray sources are more likely powered by accretion onto a 
supermassive black hole than other plausible sources of lower-luminosity 
X-ray emission, such as a population of low-mass X-ray binaries (LMXBs), 
hot halos of diffuse gas, or star formation.  

Motivated by the high AGN fraction in Abell~2104, we extended our 
survey to seven additional, low-redshift clusters of galaxies 
($z = 0.06 \rightarrow 0.31$) and found that approximately 5\% of galaxies 
more luminous than $M_R = -20$ mag host AGN more luminous than 
$L_X = 10^{41}$ \ergs\ in the broad (0.5-8 keV) X-ray band 
\citep[][hereafter Paper I]{martini06}. 
As was the case for Abell~2104, we found that most of the cluster 
galaxies with X-ray counterparts did not show obvious AGN spectral 
signatures in visible-wavelength spectra. As LMXBs or thermal emission 
from hot halos have also been observed to produce luminous X-ray emission, 
particularly from bright, early-type galaxies, we used the multiwavelength 
spectral shape of these sources to determine if they were AGN. Comparison 
with relations between X-ray and $B-$band luminosity for local early-type 
galaxies dominated by LMXBs \citep{kim04c} and hot gas 
\citep{osullivan03,sun05} showed that these AGN candidates were on order 
1 -- 3 orders of magnitude more X-ray luminous than expected from 
relations based on these other sources of X-ray emission. We therefore 
concluded that most of these sources were AGN. Only a small fraction 
remain consistent with other sources of X-ray emission and they are not 
included in the present study. 

The goal of seeking a larger sample of AGN in clusters of galaxies 
was to use the AGN population to explore the mechanisms responsible 
for fueling AGN, motivated by the similar use of the cluster environment 
to explore other aspects of galaxy evolution. Comprehensive studies of 
clusters of galaxies have shown that the cluster environment contains a 
profoundly different distribution of galaxy populations from the field. 
The morphology-density relation 
expresses the observation that lenticular and then elliptical galaxies are 
dominant in progressively richer galaxy environments \citep{dressler80}. 
The amount of current star formation in galaxies also declines toward the 
center of clusters \citep{fisher98}. Galaxies at the centers of rich 
clusters tend to be dominated by old stellar populations and have no 
active star formation, while a progressively higher fraction of poststarburst
and starburst galaxies are found toward the outskirts.  

Kinematic studies of cluster galaxies show that the population of galaxies 
with current or recent star formation have the highest velocity dispersions, 
poststarburst galaxies are intermediate, while the galaxy population dominated 
by a passive stellar population has the lowest velocity dispersion 
\citep{dressler99}. The higher velocity dispersions (and greater radial extent) 
suggest that galaxies with current or recent star formation are less virialized 
than more passive galaxies. 
These galaxies with young stellar populations may on average have 
entered the cluster more recently and have higher line-of-sight velocity 
dispersions because they remain on primarily radial orbits. 

The distribution of AGN in clusters of galaxies could provide similar 
information on the origin of AGN in clusters. In particular, their distribution 
is a valuable test of the standard paradigm for AGN fueling, namely the merger 
of two gas-rich galaxies \citep[\eg,][]{barnes92}. The low AGN fraction in 
clusters is commonly ascribed to the lower merger rate in clusters due to the 
high velocity dispersion that precludes the formation of bound pairs, in 
spite of the high galaxy density, and the lower fraction of galaxies 
with substantial reservoirs of cold gas \citep[\eg][]{giovanelli85}. 
If this picture is correct, then AGN should be more common at the 
outskirts of clusters where members are relatively rich in cold gas, as 
well as in lower velocity dispersion clusters. 
Many galaxies also enter the cluster potential in low velocity dispersion 
groups that may produce a relative increase in the AGN fraction at larger 
distances from the cluster center or associated with distinct substructure 
within the clusters. These relatively recent entrants into 
the cluster potential would also have not yet virialized and could have 
a larger velocity dispersion than the old, passively-evolving galaxies 
at the center of the cluster potential. 

In studies of lower velocity dispersion groups, \citet{shen07} found AGN in 
$\sim 7$\% of galaxies at $z \sim 0.06$, yet these AGN were only identified in 
visible-wavelength spectroscopy and were not detected in their X-ray 
({\it XMM/Newton}) observations and must have $L_X < 10^{41}$ \ergs. 
This result suggests that the X-ray and visible-wavelength properties of 
typical AGN in lower-density environments may be different from the typical 
AGN in higher-density environments. 
Previously, \citet{best05a} showed that the fraction of AGN selected by 
emission-lines from SDSS slightly decreases for galaxies with a larger 
number of luminous neighbors, yet the 
fraction of radio-selected AGN actually increases in galaxies in richer 
environments. 
This is comparable to results from the field by \citet{lehmer07}, who studied 
X-ray emission from early-type galaxies in the Extended Chandra Deep Field 
South and found an average AGN fraction with hard X-ray luminosity above 
$10^{41}$ \ergs\ consistent with our measurement in clusters. For the broad 
X-ray band they find: $f_A(M_R<-20;L_X>10^{41}) \sim 7$\% and 
$f_A(M_R<-20;L_X>10^{42}) = 2$\% (B. Lehmer 2006, private communication). 
These fractions are remarkably similar to the mean AGN fractions we measure 
in clusters and may suggest the X-ray luminous AGN fraction is not a strong 
function of environment, similar to the result of \citet{best05a} for radio 
luminous AGN, although this does not explain the absence of X-ray luminous 
AGN in the group study of \citet{shen07}. 

AGN in clusters of galaxies have also garnered significant recent 
interest as an explanation of the absence of substantial cold gas at the 
centers of many clusters. Simple radiative cooling models predict that 
the intracluster medium (ICM) in the cores of many clusters should cool in 
less than a Hubble time, yet the predicted substantial reservoirs of cold 
gas are not observed. The presence of powerful radio galaxies at the 
centers of most predicted 'cool-core' clusters \citep[e.g.,][]{burns90} 
suggests that while the cooling gas may provide fuel for the AGN, the AGN may 
also be reheating the cool gas. 
The substantial cavities in the hot ICM 
coincident with the lobes of these radio galaxies demonstrates that the AGN 
inject copious amounts of energy into the ICM \citep{mcnamara00,fabian00}. 
\citet{birzan04} have shown that the energy 
necessary to reheat the ICM and prevent cooling is approximately consistent 
with the amount of energy necessary to create the large cavities 
observed in the ICM, although the mechanism by which the highly collimated 
AGN jets uniformly heat the core ICM is still under active investigation. 
One possible resolution to the problem of uniform heating is multiple 
AGN in the cluster core ($\sim 100$ kpc). \citet{nusser06} recently 
showed that gas cooling at the centers of clusters could condense onto 
any galaxies within the core, fuel accretion onto the supermassive black holes, 
and produce more distributed heating of the ICM. A simple test of this 
scenario is to search for multiple or off-center AGN in the cores of 
clusters. It is also important to quantify the population and evolution 
of radio-bright AGN in the cores of clusters for future experiments that 
employ the Sunyaev-Zel'dovich effect to identify clusters because these 
sources could be a significant contaminant \citep[e.g.,][]{coble07}. 

In the present paper we employ the other cluster members identified 
in our multiwavelength survey to derive the spatial and kinematic 
distribution of the cluster AGN relative to the cluster galaxy population. 
We also use these data to derive velocity dispersions and study the 
amount of star formation present in these clusters to determine if 
any global cluster properties correlate with the cluster AGN fraction. 
In the next section we provide a brief summary of 
the observations described in Paper I, followed by a derivation of the 
velocity dispersion, membership, and visible-wavelength properties of the 
cluster galaxies in \S~\ref{sec:clusters}. The distribution of the 
AGN relative to other cluster members is described in \S~\ref{sec:dist} and 
the relation between AGN fraction and the properties of the 
cluster is discussed in \S~\ref{sec:relation}. We present our conclusions in 
the last section. Throughout this paper we assume that the cosmological 
parameters are: ($\Omega_M, \Omega_\Lambda, h$) = (0.3, 0.7, 0.7) where 
$H_0 = 100h$ \kms Mpc$^{-1}$. 

\section{Observations} \label{sec:obs} 

\begin{deluxetable*}{lccccccc}
\tablecolumns{8}
\tablewidth{7.0truein}
\tabletypesize{\scriptsize} 
\tablecaption{Cluster Properties\label{tbl:clusters}}
\tablehead{
\colhead{Cluster} &
\colhead{$\alpha_c$} &
\colhead{$\delta_c$} &
\colhead{$z$} &
\colhead{$z_1,z_2$} &
\colhead{$\sigma$} & 
\colhead{$r_{200}$} & 
\colhead{P$_\delta$} \\
\colhead{(1)} &
\colhead{(2)} &
\colhead{(3)} &
\colhead{(4)} &
\colhead{(5)} &
\colhead{(6)} & 
\colhead{(7)} & 
\colhead{(8)} \\  
}
\startdata
Abell 3125    & 03:27:17.9 & -53:29:37 & 0.0616 & 0.0530,0.0700 & 475   (94) & 1.14 & $<0.001$ \\
Abell 3128    & 03:30:43.8 & -52:31:30 & 0.0595 & 0.0435,0.0755 & 906   (74) & 2.18 & $<0.001$ \\
Abell 644     & 08:17:25.6 & -07:30:45 & 0.0701 & 0.0531,0.0871 & 952  (382) & 2.28 & 0.236 \\
Abell 2104    & 15:40:07.9 & -03:18:16 & 0.1544 & 0.1304,0.1783 & 1242 (194) & 2.85 & 0.203 \\
Abell 1689    & 13:11:29.5 & -01:20:28 & 0.1867 & 0.1392,0.2343 & 2402 (357) & 5.41 & 0.994 \\
Abell 2163    & 16:15:49.0 & -06:08:41 & 0.2007 & 0.1731,0.2236 & 1381 (324) & 3.09 & 0.283 \\
MS1008        & 10:10:32.4 & -12:39:53 & 0.3068 & 0.2921,0.3215 & 1127 (153) & 2.38 & 0.722 \\
AC114         & 22:58:48.4 & -34:48:08 & 0.3148 & 0.2884,0.3412 & 2025 (217) & 4.25 & 0.156 \\
\enddata
\tablecomments{
Cluster sample and properties derived from the present study. Columns are: (1) 
Cluster name; (2 and 3) RA and DEC of the cluster center for epoch J2000; (4) 
redshift; (5) redshift range of cluster members; (6) velocity dispersion and 
uncertainty; (7) $r_{200}$ in Mpc; (8) probability that the observed 
substructure is consistent with a random redistribution of the cluster radial 
velocities.  The derivation of these quantities is discussed in 
Sections~\ref{sec:clusters} and \ref{sec:dist}. 
}
\end{deluxetable*}

\begin{figure*}
\figurenum{1} 
\plotone{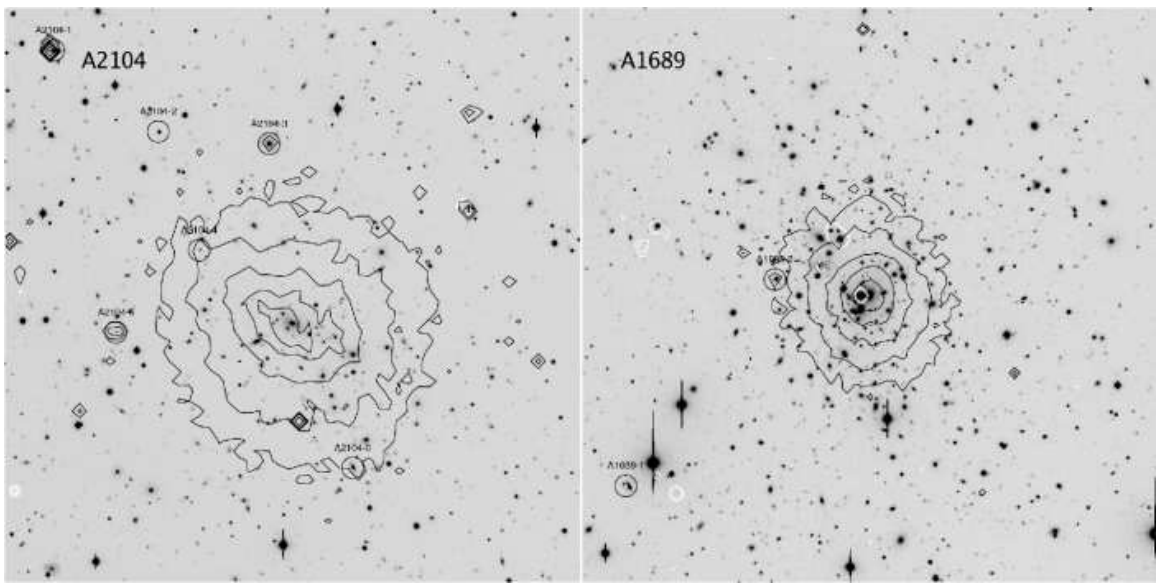}
\caption{
\label{fig:image1} 
R-band images of Abell 2104 ($z=0.154$) and Abell 1689 ($z=0.187$) with 
\chandra\ X-ray ({\it black contours}) and FIRST radio ({\it white contours}) 
data.  Each image is $7' \times 7'$ on a side, North is up and East is to the 
left. These images only show the central region of our data where the X-ray 
emission in prominent. 
}
\end{figure*}

The goal of this survey was to identify AGN in clusters of galaxies. This 
was accomplished with X-ray and visible-wavelength images and spectroscopy 
of the eight clusters of galaxies listed in Table~\ref{tbl:clusters}. As AGN 
are quite luminous at X-ray wavelengths, 
AGN candidates were identified as galaxies with X-ray counterparts. These 
galaxies were then observed spectroscopically to determine if they were 
members of a given cluster. This X-ray selection technique was employed 
for two reasons. First, AGN are relatively rare in clusters of galaxies 
and it is more efficient to obtain complete spectroscopic observations of all 
AGN candidates than to 
spectroscopically observe all bright galaxies because the surface density 
of candidate cluster members with X-ray counterparts is substantially 
lower than the surface density of all candidate cluster members. Secondly, 
X-ray selection is a relatively unbiased method of identifying AGN over a 
wide range of redshifts. Most AGN are low-luminosity (e.g., $L_X < 10^{43}$ 
\ergs) and these AGN are difficult to identify spectroscopically at even 
low-redshift (\eg, $z \sim 0.1$) due to dilution by host galaxy starlight 
and/or obscuration. X-ray emission from an AGN has substantially higher 
contrast over other sources of X-ray emission to much lower accretion power, 
particularly when 
the spectroscopic aperture includes a substantial fraction of the total 
emission of the galaxy. X-ray emission is also much less affected by 
obscuration than visible-wavelength emission. 

\begin{figure*}
\figurenum{2} 
\plotone{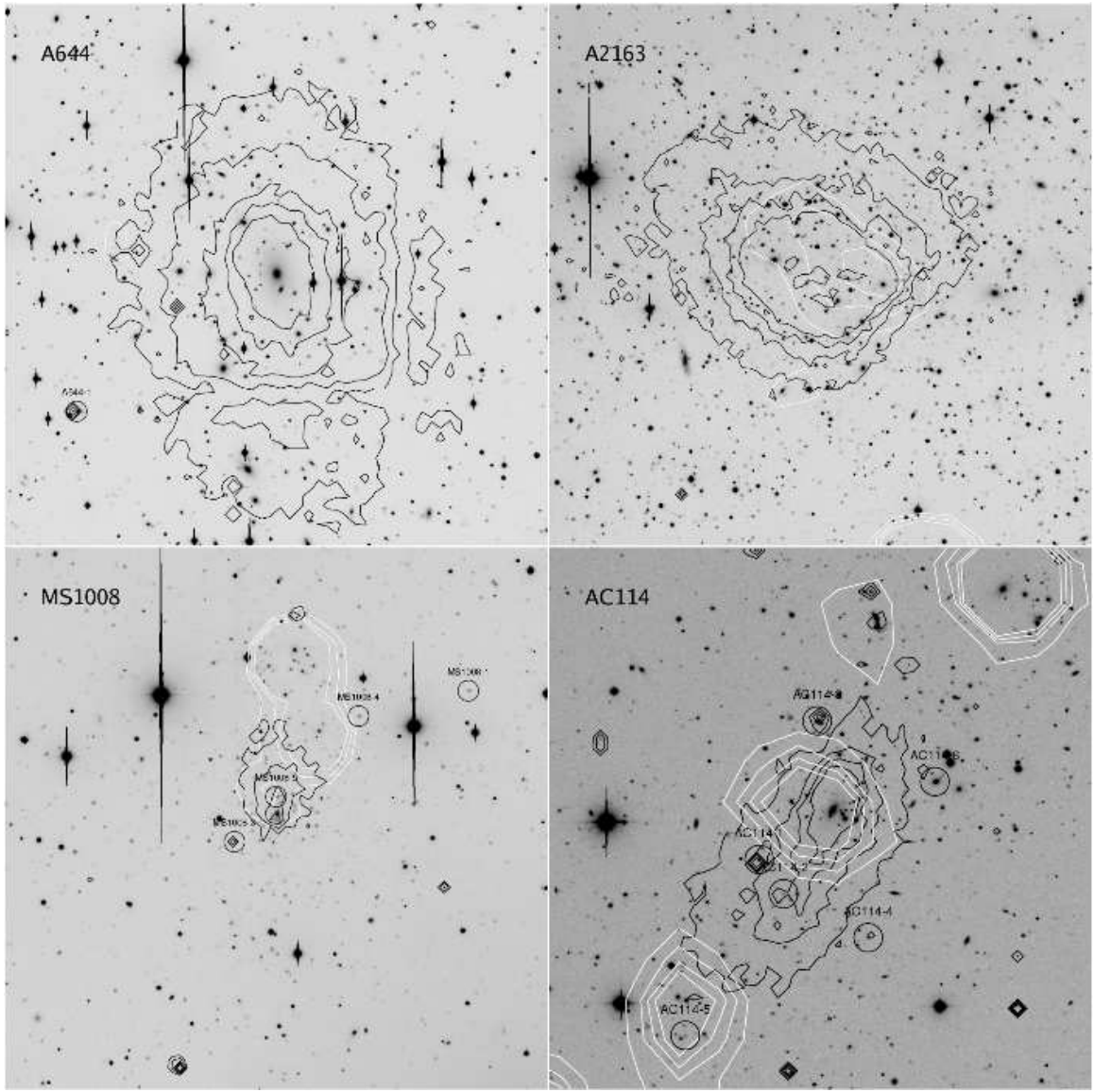}
\caption{
\label{fig:image2} 
R-band images of Abell 644 ({\it upper left}, $z=0.070$), Abell 2163 
({\it upper right}, $z=0.201$), MS1008 ({\it lower left}, $z = 0.307$), 
and AC114 ({\it lower right}, $z = 0.315$) along with \chandra\ X-ray ({\it 
black contours}) and NVSS radio ({\it white contours}) data. 
Each image is $7' \times 7'$ on a side, North is up and East is to the 
left. As in Figure~\ref{fig:image1}, these images only show the central 
region of the cluster where the X-ray emission is prominent. 
}
\end{figure*}

The observing strategy was therefore to compare visible-wavelength and X-ray 
images of each cluster to identify potential cluster AGN for follow-up 
spectroscopy. 
As discussed in greater detail in Paper I, the X-ray observations are 
archival \chandra\ observations of low-redshift clusters with sufficient 
sensitivity to identify low-luminosity AGN. We obtained visible-wavelength 
observations of the clusters with the 2.5m du Pont telescope at Las 
Campanas Observatory with either the WFCCD or Tek5 CCD cameras. $R-$band 
images of the centers of six of these clusters are shown in 
Figures~\ref{fig:image1} and \ref{fig:image2} along with X-ray and radio 
contours. Comparable figures for A3125 and A3128 are presented in 
\citet{rose02}. 
All X-ray counterparts brighter than $R\sim23$ mag were then selected 
for multislit spectroscopic observations. As the surface density of X-ray 
counterparts was substantially less than the potential packing density of 
spectroscopic slits on a given mask, additional bright galaxies were observed 
to determine the spectroscopic properties of cluster members without X-ray 
emission, measure the fraction of cluster members with X-ray emission, and in 
several cases derive the velocity dispersion of the cluster for the first 
time. 

Table~\ref{tbl:catalog} presents a catalog of all of our successful 
spectroscopic observations. As noted above, these targets were selected 
through a variety of algorithms. First, the highest priority 
was assigned to all X-ray sources brighter than $R\sim23$ mag. Additional 
candidate cluster members were then targeted for spectroscopy based on 
brightness, color, and availability of a spectroscopic slit. 
We successfully measured spectroscopic redshifts for all sources brighter 
than a rest-frame absolute magnitude of $M_R = -20$ mag at each 
cluster redshift. For the nearby clusters, the apparent magnitude limits 
extend substantially fainter. In general, we are complete to 
$R = 21$ mag for all of these fields and to $R = 22$ mag for all sources 
selected as X-ray counterparts. In addition, there are many fainter 
sources with emission-line redshifts. The exception is the Abell~2163 field, 
where the 90\% completeness for spectroscopy is approximately $R = 20.5$ mag 
for both selection criteria. With the exception of Abell~644 and Abell~2163, 
there are literature redshift data for all of these fields. We observed a 
number of these sources in each field and used these data to determine the 
typical redshift uncertainty of our observations is less than 
$\sigma_z = 0.0005$. 

\begin{deluxetable*}{ccccccccccl}
\tablecolumns{11}
\tabletypesize{\scriptsize} 
\tablecaption{Spectroscopic Catalog\label{tbl:catalog}}
\tablehead{
\colhead{RA} & 
\colhead{DEC} & 
\colhead{$z$} &
\colhead{Template} &
\colhead{Mask} &
\colhead{Select} &
\colhead{$R$} &
\colhead{$B-R$} &
\colhead{$V-R$} &
\colhead{$R-I$} &
\colhead{Lit ID} \\
\colhead{(1)} &
\colhead{(2)} &
\colhead{(3)} &
\colhead{(4)} &
\colhead{(5)} &
\colhead{(6)} &
\colhead{(7)} &
\colhead{(8)} &
\colhead{(9)} &
\colhead{(10)} & 
\colhead{(11)} \\ 
}
\startdata
3:27:45.35 & -53:24:02.6 &      0.0599 & A & a3125a &   P &      16.89 (0.03) &   1.65 (0.05) &   0.59 (0.05) &            &      \\
3:27:50.22 & -53:24:54.7 &      0.5402 & E & a3125a &   P &      20.62 (0.03) &   1.72 (0.10) &   0.94 (0.06) &            &      \\
3:27:33.96 & -53:23:52.1 &      0.0626 & A & a3125a &   P &      17.37 (0.03) &   1.66 (0.05) &   0.59 (0.05) &            &      \\
3:27:22.03 & -53:25:57.6 &      0.3950 & E & a3125a &   P &      21.14 (0.04) &               &   0.87 (0.11) &            &      \\
3:27:52.60 & -53:24:08.4 &      0.0614 & A & a3125a &   P &      15.93 (0.03) &   1.77 (0.05) &   0.66 (0.05) &            &    2MASX J03275262-5324079\\
3:27:46.86 & -53:22:58.7 &      0.2581 & E & a3125a &   P &      20.16 (0.03) &   1.62 (0.06) &   0.64 (0.05) &            &      \\
3:27:46.88 & -53:22:08.8 &      0.3937 & E & a3125a &   P &      20.29 (0.03) &   1.66 (0.06) &   0.69 (0.05) &            &      \\
3:27:56.19 & -53:23:18.4 &      0.4374 & E & a3125a &   P &      20.88 (0.03) &   1.40 (0.08) &   0.68 (0.05) &            &      \\
3:27:38.88 & -53:26:04.5 &      0.597  & Q & a3125a &   X &      19.91 (0.03) &   0.35 (0.05) &   0.14 (0.05) &            &      \\
3:27:54.32 & -53:21:51.1 &      0.84   & Q & a3125a &   X &      20.80 (0.03) &   0.64 (0.05) &   0.28 (0.05) &            &      \\
3:26:55.66 & -53:31:58.8 &      0.3192 & A & a3125b &   P &      19.18 (0.03) &   2.71 (0.07) &   0.97 (0.05) &            &      \\
3:26:54.48 & -53:31:47.9 &      0.3101 & E & a3125b &   P &                   &               &               &            &      \\
3:27:00.96 & -53:31:36.6 &      0.0630 & E & a3125b &   P &      19.91 (0.03) &   1.09 (0.05) &   0.38 (0.05) &            &      \\
3:27:30.01 & -53:31:04.4 &      0.2613 & A & a3125b &   P &      19.72 (0.03) &   1.55 (0.05) &   0.56 (0.05) &            &      \\
3:27:22.70 & -53:30:32.9 &      0.5374 & A & a3125b &   P &      20.75 (0.03) &   1.29 (0.07) &   0.84 (0.06) &            &      \\
\enddata
\tablecomments{
Spectroscopic catalog of all sources with successful redshift 
measurements. Columns are: (1 and 2) object RA and DEC in J2000; 
(3) redshift; (4) best template match as emission-line galaxy (E), 
absorption-line galaxy (A), quasar (Q), or Galactic star (S); 
(5) mask identification; (6) target selection as either an 
X-ray source or from R-band photometry; (7) R-band magnitude and uncertainty; 
(8 to 10) $B-R$, $V-R$, and $R-I$ color and uncertainty; (11) literature 
identification. The complete version of this table is in the electronic 
edition of the Journal. The printed edition contains only a sample.
}
\end{deluxetable*}

\section{Cluster Properties} \label{sec:clusters} 

\begin{figure*}
\figurenum{3} 
\plotone{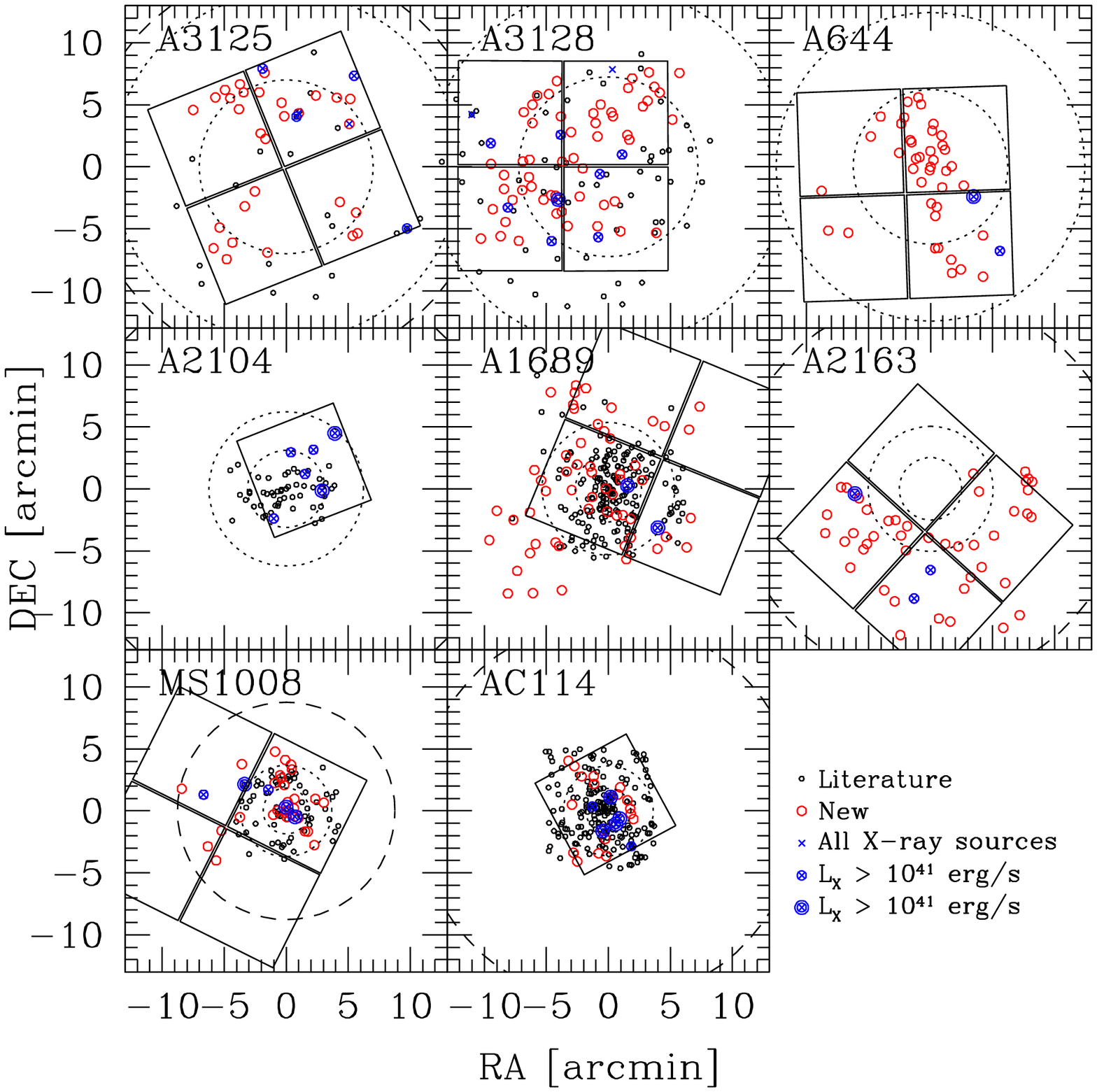} 
\caption{
\label{fig:pos} 
Positions of cluster members relative to the centers listed in 
Table~\ref{tbl:clusters}. Members are marked with different symbols if 
their positions are from the literature ({\it small, black circles}), 
identified in the present study ({\it large, red circles}), and 
X-ray sources ({\it blue crosses}). Also shown are the field of view of the 
\chandra\ observations ({\it solid, black boxes}) and circles of 
radius 0.5 Mpc ({\it smaller, dotted circle}), 1 Mpc ({\it larger, dotted 
circle}), and $r_{200}$ ({\it large, dashed circle}), if the later radius 
falls within the field of view shown. 
The axes on all panels are offsets in arcminutes from the cluster center. 
}
\end{figure*}

We use the membership data and the $R-$band images and X-ray contours shown in 
Figures~\ref{fig:image1} and \ref{fig:image2} to either derive new centers 
for each cluster or confirm a previously-reported center from the literature. 
These cluster centers are listed in Table~\ref{tbl:clusters}. 
For most of the clusters, either a bright cluster member or the brightest 
cluster galaxy (BCG) is coincident with the approximate peak of the 
ICM. In these cases we adopt the coordinates of this 
galaxy for the cluster center, although for Abell~2163 the BCG candidate 
was not observed spectroscopically. Only Abell~3125 and Abell~3128 do 
not have obvious, bright cluster galaxies at their centers. Abell~3125 in 
fact does not have a diffuse ICM, while the ICM for Abell~3128
is double-peaked. This pair of merging clusters have been extensively 
discussed and modeled by \citet{rose02}, and we adopt their position for the 
center of Abell~3128, which is approximately midway between the peaks in the 
extended X-ray emission. For Abell~3125 we simply adopt the mean position 
of all of the confirmed members, which is within an arcminute of the 
position reported by \citet{abell89}. Figure~\ref{fig:pos} shows the 
distribution of cluster members relative to the adopted centers along with the 
area subtended by the \chandra\ field of view. 

\subsection{Velocity Dispersions and Membership } 

\begin{figure*}
\figurenum{4} 
\plotone{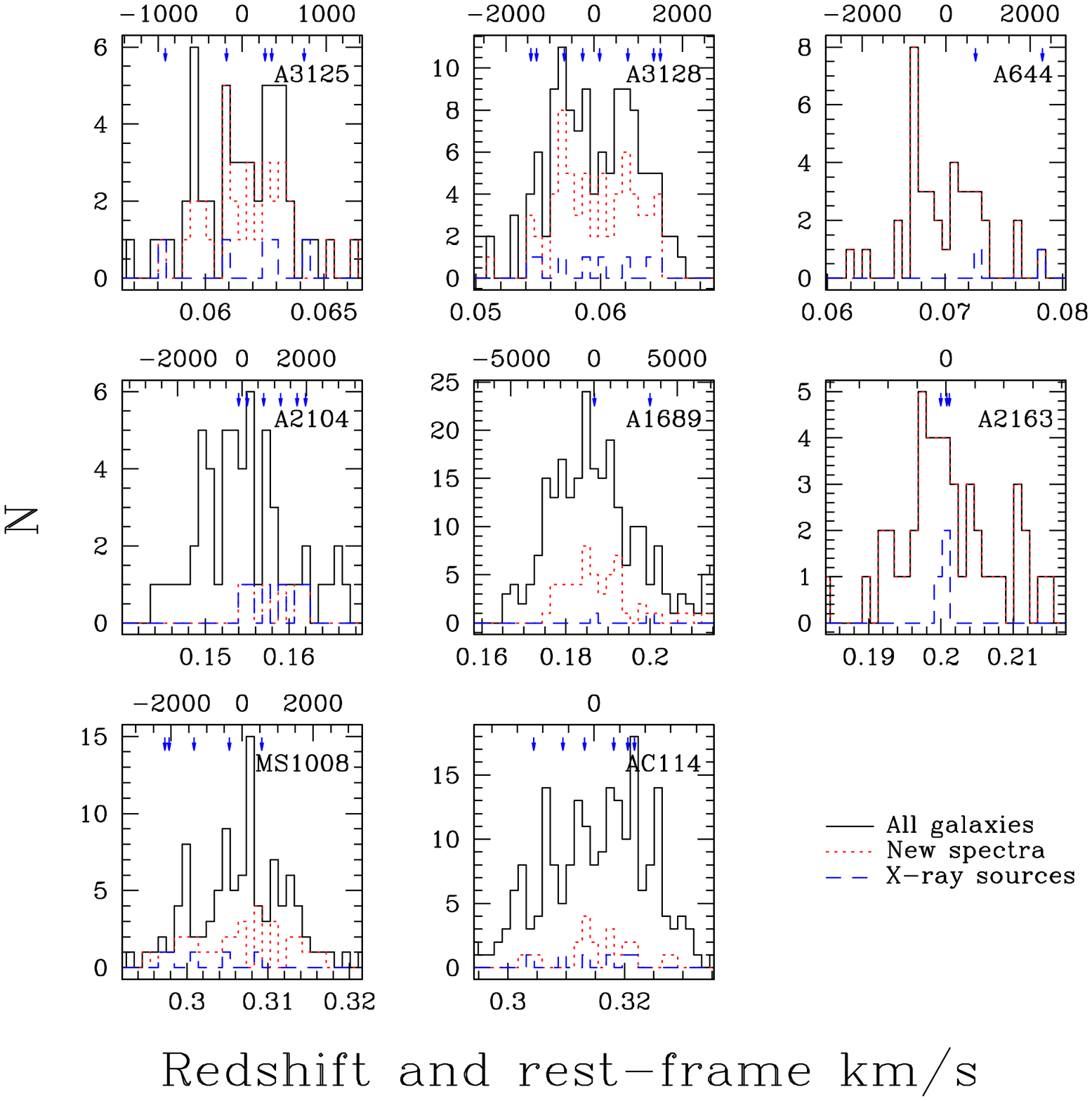} 
\caption{
\label{fig:sigma} 
Velocity distributions for each cluster. Each panel displays the distribution 
of all galaxies within $3\sigma$ of the cluster mean redshift ({\it black, 
solid line}), the subset of new members ({\it red, dotted line}), and the 
X-ray sources ({\it blue, dashed line}) for a cluster field. The redshifts 
of individual X-ray sources are also marked ({\it blue arrows}). The 
bottom abscissa plots the observed redshift, while the top abscissa 
plots the rest-frame velocity offset from the cluster mean. 
}
\end{figure*}

We have used the spectroscopic data presented in Table~\ref{tbl:catalog} 
and literature membership data to determine the redshift, velocity 
dispersion, and redshift limits of each of these clusters. These values are 
presented in Table~\ref{tbl:clusters}. We calculate these values with an 
implementation of the center and scale estimators described in 
\citet{beers90}. For each cluster we combine our redshift measurements 
and any from the literature and calculate an approximate redshift and 
velocity dispersion using the biweight estimator for center and scale. 
We then removed galaxies more than $5\sigma$ from the center and iterated to 
determine the mean redshift and velocity dispersion of each cluster. The 
adopted redshift range is $\pm 5\sigma$ of the cluster mean. 
The uncertainty in the velocity dispersion is the 90\% confidence limit 
calculated with the jackknife of the biweight estimator. 
Figure~\ref{fig:sigma} shows the galaxy velocity distribution for 
each cluster over the range $\pm 3\sigma$. 

As noted above, there are literature measurements of multiple cluster 
members for all of these clusters except Abell~644 and Abell~2163. 
\citet{rose02} performed a detailed study of the merging clusters Abell~3125 
and Abell~3128 with spectroscopic observations over a 2 degree field and mapped 
out the complex velocity structure of this cluster pair. Our velocity 
measurements are broadly consistent with their results, although given the 
kinematic complexity of this system the derived velocity dispersions are 
unlikely to be a good measure of the cluster mass. As our observations have a 
substantially smaller field of view than the \citet{rose02} study and map only 
a small fraction of the physical extent of these clusters, the present 
observations are not as well suited to map the velocity structure in detail. 
Our measurements for Abell~2104, Abell~1689, MS1008, and AC~114 are in good 
agreement with previous values. This is least surprising for Abell~2104, 
since we only have new redshifts for six X-ray counterparts to add to the study 
of \citet{liang00}. AC~114 has the largest difference from the literature 
value. We measure a 15\% larger velocity dispersion than the earlier estimate 
by \citet{couch87}, although a substantial number of new members have been 
identified since that work. Our value for Abell~1689 is comparable to the 
early measurement of \citet{teague90} and larger than the recent value of 
\citet{czoske04}, who measure $\sigma \approx 2100$ \kms\ based on a 
larger sample of over 500 (unpublished) members; however, the difference 
between these values has no impact on our results below.

  \subsection{Galaxy Colors and Spectral Properties} \label{sec:props} 

\begin{figure*}
\figurenum{5} 
\plotone{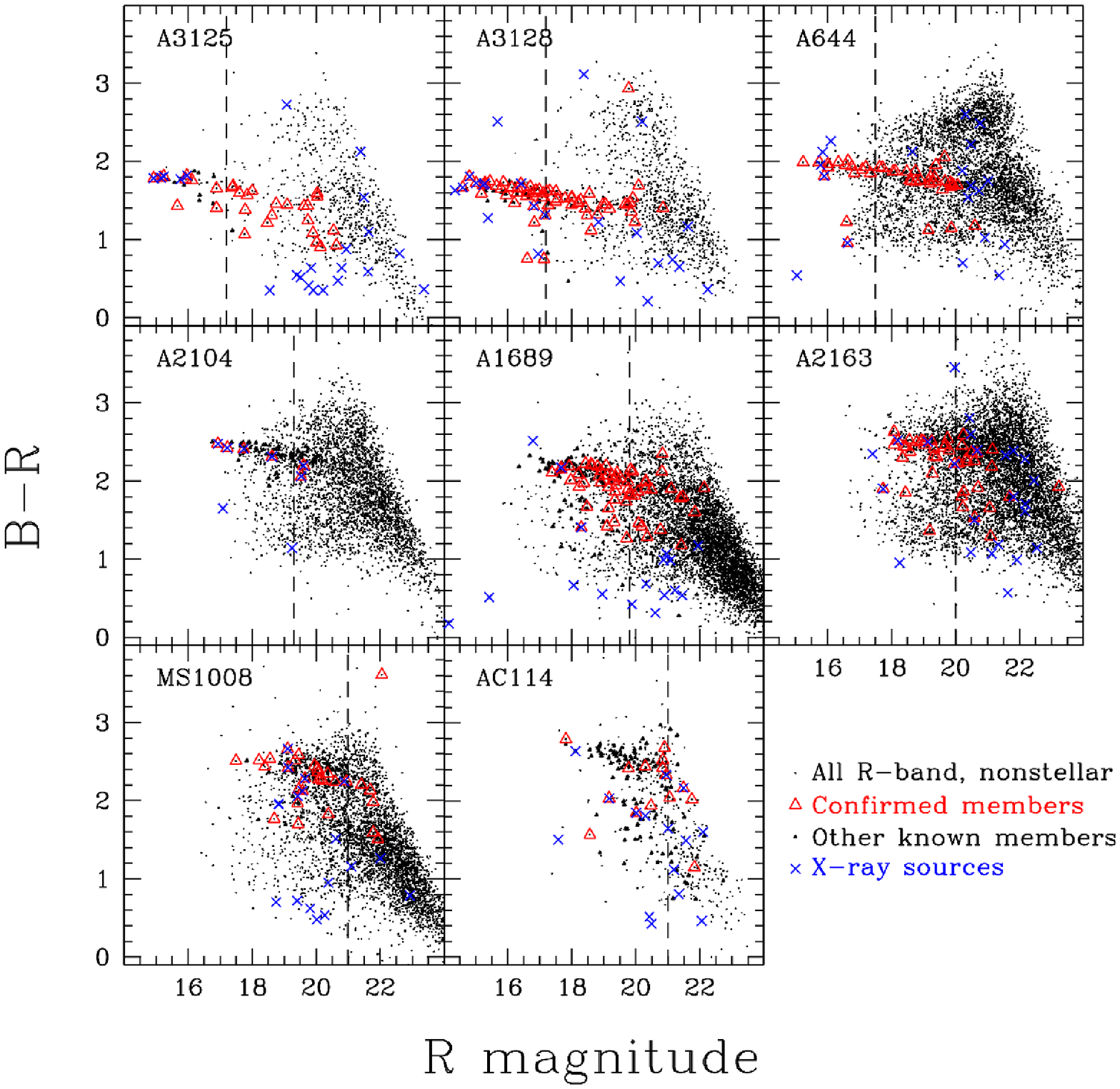} 
\caption{
\label{fig:cmd} 
Color-magnitude diagrams for the eight cluster fields. All sources in 
our $R-$band catalog with measured $B-R$ colors and $R-$band stellarity 
less than 0.9 are shown. Confirmed members from Table~\ref{tbl:catalog} 
trace the color-magnitude relation of the passively-evolved galaxies in 
each cluster ({\it red, open triangles}). Other members from the literature 
({\it small triangles}) and X-ray sources ({\it blue crosses}) are also 
marked. The vertical, dashed line marks the apparent magnitude that corresponds 
to a rest-frame $M_R = -20$ mag for each cluster. 
}
\end{figure*}

In addition to changes in the mix of galaxy populations within clusters, 
such as the morphology-density relation \citep{dressler80}, the mix of 
galaxy populations can also vary between clusters. One clear example of 
this is the Butcher-Oemler effect \citep{butcher78}, the tendency for 
clusters to have larger blue galaxy fractions at higher redshift. Specifically, 
\citet{butcher78} quantified the blue galaxy fraction in a cluster as the 
fraction of galaxies more than 0.2 mag bluer in rest-frame $B-V$ color. 
The color-magnitude diagrams presented in Figure~\ref{fig:cmd} indicate that 
the confirmed cluster members ({\it triangles}) have a substantial variation 
in observed $B-R$ color. We calculate the Butcher-Oemler (BO) galaxy fraction 
for each cluster as the fraction of all known members more luminous than 
$M_R < -20$ mag in our photometric data set. The number of BO galaxies and 
BO fraction are listed in Table~\ref{tbl:popfrac}. The highest fractions 
are measured for clusters with substantial starforming galaxy populations 
(e.g., A1689, AC114). While the BO fraction is likely sensitive to potential 
biases such as the fraction of the cluster surveyed (e.g., fraction of the 
virial radius) and the limiting magnitude of the spectroscopy, the relative 
fractions provide a measure of the relative star formation activity from 
cluster to cluster in this sample. The difference between our measured BO 
fraction of 18\% for AC114 and the 29\% measured by \citet{couch87} 
provides a reasonable gauge of how sensitive this quantity is to measurement 
details. 

We have also used our spectroscopic data for all of the clusters except 
Abell~2104 to determine the emission-line galaxy fraction for galaxies 
more luminous than $M_R < -20$ mag. This provides a second measure of the 
frequency of star formation in cluster members, although this quantity is 
less reliable than the BO fraction because emission-lines may be 
more difficult to detect in cluster members at higher-redshift because the 
typical signal-to-noise ratio of the spectroscopy is lower. We simply 
classify any galaxy with an emission line as an emission-line galaxy, 
regardless of the strength of the line. In the vast majority of cases, the 
observed emission-line is the [OII] $\lambda3727$ doublet and nearly all of the 
emission-line galaxies appear to be starforming galaxies. The few exceptions 
are the small number of X-ray-identified AGN with visible-wavelength 
emission-line signatures. We do not spectroscopically identify any AGN that 
do not have an X-ray counterpart. Both the BO fraction and emission-line 
galaxy fraction for these clusters are compared with the AGN fraction 
in Section~\ref{sec:relation} below. 

\begin{deluxetable}{lcccc}
\tablecolumns{5}
\tablewidth{3.0truein}
\tabletypesize{\scriptsize}
\tablecaption{Cluster Galaxy Population Fractions \label{tbl:popfrac}}
\tablehead{
\colhead{Cluster} &
\colhead{$N_{BO}$} &
\colhead{$f_{BO}$} &
\colhead{$N_{e}$} &
\colhead{$f_{e}$} \\
\colhead{(1)} &
\colhead{(2)} &
\colhead{(3)} &
\colhead{(4)} &
\colhead{(5)} \\ 
}
\startdata
A3125   &  2 & 0.11 & 4  & 0.44  \\
A3128   &  5 & 0.07 & 7  & 0.21  \\
A644    &  2 & 0.13 & 1  & 0.06  \\
A2104   &  3 & 0.09 & ... & ... \\
A1689   & 31 & 0.25 & 8  & 0.24  \\
A2163   &  3 & 0.12 & 3  & 0.11  \\
MS1008  & 18 & 0.31 & 4  & 0.16  \\
AC114   & 25 & 0.26 & 2  & 0.18  \\
\enddata
\tablecomments{
The fraction of Butcher-Oemler and emission-line galaxies in each cluster that 
are brighter than a fixed rest-frame absolute magnitude of $M_R = -20$ mag. 
For each cluster listed in column~1 we list the total number of Butcher-Oemler 
galaxies that are known cluster members in column~2 and their fraction of the 
total cluster galaxy population above that luminosity in column~3. We 
present the total number of emission-line galaxies and the emission-line galaxy 
fraction in columns~4 and 5, although this quantity is only calculated 
from galaxies we observed spectroscopically. We do not have sufficient data 
to calculate this quantity for A2104. 
See \S~\ref{sec:props} for further details.
}
\end{deluxetable}

\section{Distribution of Cluster AGN}  \label{sec:dist} 

\begin{deluxetable}{llc}
\tablecolumns{3}
\tablewidth{3.0truein}
\tabletypesize{\scriptsize} 
\tablecaption{KS Test Results \label{tbl:ks}}
\tablehead{
\colhead{Distribution Type} &
\colhead{Sample Limits} &
\colhead{KS} \\ 
\colhead{(1)} &
\colhead{(2)} &
\colhead{(3)} \\
}
\startdata
Velocity distribution 	& $L_X > 10^{41}$		& 0.683 \\ 
			& $L_X > 10^{42}$		& 0.955 \\ 
			& $L_X > 10^{41}-$A3125/8 	& 0.388 \\ 
			& $L_X > 10^{42}-$A3125/8 	& 0.967 \\ 
Radial [Mpc]            & $L_X > 10^{41}$		& 0.743 \\ 
			& $L_X > 10^{42}$		& 0.069 \\ 
			& $L_X > 10^{41}-$A3125/8 	& 0.950 \\ 
			& $L_X > 10^{42}-$A3125/8 	& 0.128 \\ 
Radial [$r_{200}$]	& $L_X > 10^{41}$		& 0.989 \\ 
			& $L_X > 10^{42}$		& 0.028 \\ 
			& $L_X > 10^{41}-$A3125/8 	& 0.513 \\ 
			& $L_X > 10^{42}-$A3125/8 	& 0.051 \\ 
Substructure		& $L_X > 10^{41}$		& 0.616 \\ 
			& $L_X > 10^{42}$		& 0.606 \\ 
			& $L_X > 10^{41}-$A3125/8 	& 0.302 \\ 
			& $L_X > 10^{42}-$A3125/8 	& 0.649 \\ 
\enddata
\tablecomments{
Results of the Kolmogorov-Smirnoff tests. Columns (1) and (2) describe the 
input sample of X-ray sources; in all cases the comparison sample comprises 
the cluster members without X-ray emission. Column (3) lists the probability 
that the X-ray sources are drawn from a different distribution. Sample 
limits marked $-$A3125/8 did not include these two clusters. 
See \S~\ref{sec:dist} for further details. 
}
\end{deluxetable}

Major mergers between gas-rich galaxies remain the standard paradigm for 
triggering high-luminosity quasars, although there is little direct evidence 
that such spectacular events are responsible for triggering lower-luminosity 
AGN as well \citep[\eg,][]{derobertis98,schmitt01}. 
However, in the standard picture of merger-driven fueling, 
the AGN luminosity diminishes gradually over up to a Gyr 
\citep[\eg,][]{hopkins05}. Little evidence of a past merger may therefore 
be present during a more extended, low-luminosity phase. 

As outlined above, major mergers between gas-rich galaxies should be 
substantially rarer in clusters of galaxies than in the field because the 
velocity dispersion is too high for major mergers and few galaxies are gas 
rich. The most favorable region for major mergers to still occur in clusters 
is in the outskirts, which contain many galaxies falling into the cluster for 
the first time. These galaxies are primarily spirals and therefore are richer 
in cold gas. They are also often bound in low-velocity groups with velocity 
dispersions of a few hundred kilometers per second and consequently have a 
higher merger probability.

Previous studies have shown that there is more activity in the outskirts 
of clusters, such as a larger fraction of emission-line and poststarburst 
galaxies \citep[\eg,][]{couch87,fabricant91,fisher98}, although these 
studies did not specifically address the AGN distribution. 
In a study of 10 low-redshift clusters, \citet{dressler99} found that 
emission-line galaxies had a higher velocity dispersion than the more passive 
galaxies. The higher line-of-sight velocity dispersion implies that the 
emission-line galaxies are not yet virialized and remain on primarily radial 
orbits in the cluster potential. 

Clusters of galaxies offer a unique opportunity to test if low-luminosity 
AGN are primarily the result of the major mergers of gas-rich galaxies 
because these mergers are most likely to occur in the infalling population 
and the cluster crossing time is comparable to the predicted lifetime of the 
low-luminosity phase \citep[\eg,][]{martini04c}. 
Low-luminosity AGN triggered in infalling galaxies 
will therefore remain on primarily radial orbits after a Gyr, long after 
evidence of their violent past has faded. 
In the next subsections we investigate the kinematic and radial 
distribution of the cluster AGN to determine if they may have recently 
entered the cluster potential. 

  \subsection{Velocity Distribution}  \label{sec:veldist} 

In Figure~\ref{fig:sigma} we show the galaxy velocity distribution for 
each cluster and mark the locations of all of the X-ray sources with 
arrows and a dashed histogram. While there are a number of X-ray sources 
several standard deviations from the mean cluster redshift, there are an 
insufficient number of sources per cluster to state if these outliers 
represent a significant fraction of the population. 

\begin{figure}
\figurenum{6} 
\plotone{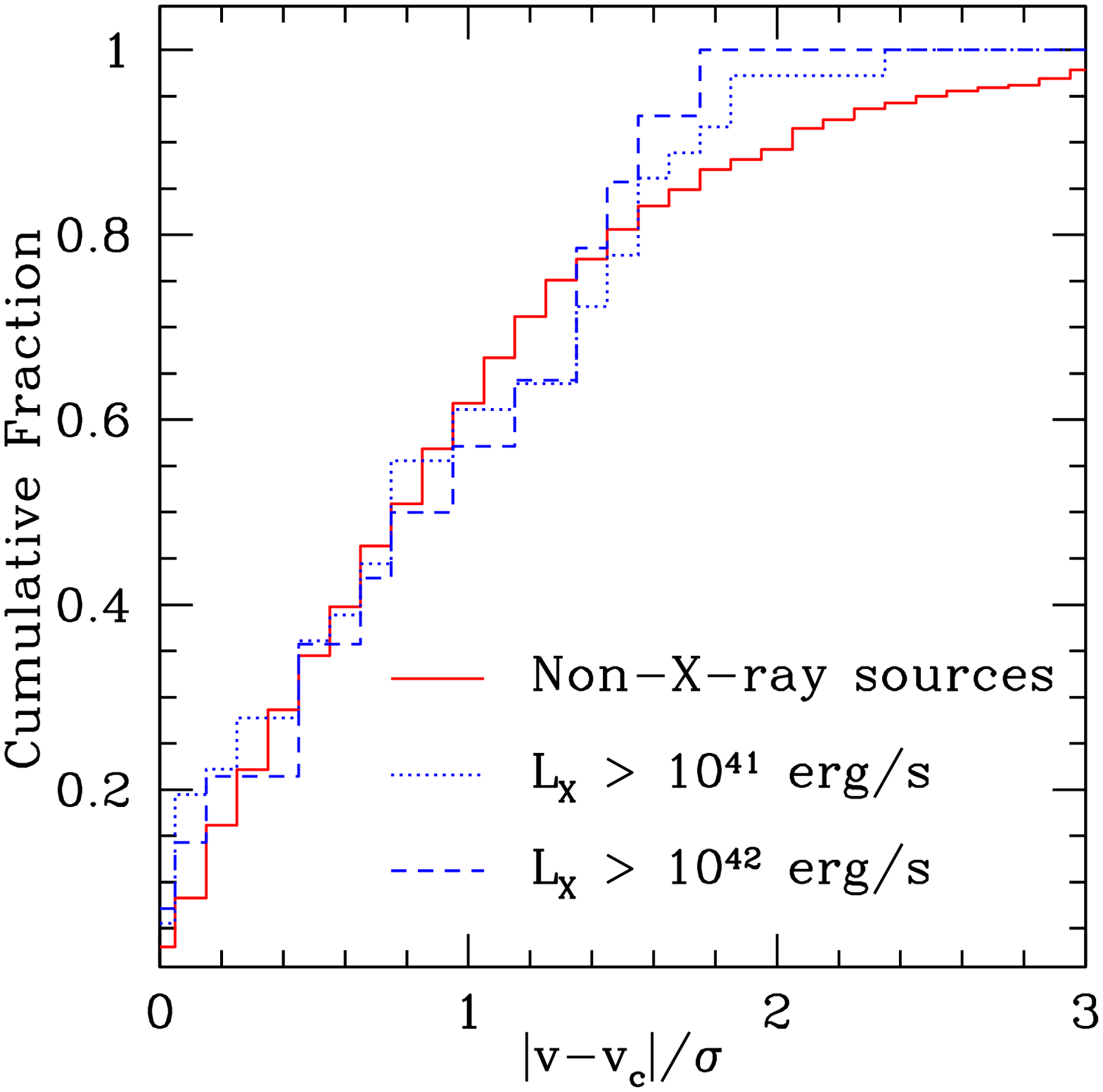} 
\caption{
Cumulative, normalized velocity distribution of cluster X-ray sources 
relative to each cluster's mean redshift. The distribution of the AGN more 
luminous than $L_X > 10^{41}$ \ergs ({\it blue, dotted line}), more luminous 
than $L_X > 10^{42}$ \ergs ({\it blue, dashed line}), and the inactive 
galaxies ({\it red, solid line}) are statistically indistinguishable. 
The absolute value of the velocity 
offset of each galaxy from the cluster mean has been normalized to the 
cluster's velocity dispersion. See \S\ref{sec:veldist} for further 
details. 
\label{fig:veldist} 
}
\end{figure}

To increase the AGN sample size for statistical analysis, we have combined the 
velocity distributions for all eight clusters. 
Because there is a substantial variation in the number of galaxies, number 
of X-ray sources, and velocity dispersions, we combined these clusters 
by first calculating the absolute value of the velocity offset of each 
galaxy from the cluster mean with respect to the cluster velocity dispersion. 
All eight clusters were then summed in these normalized coordinates and 
in Figure~\ref{fig:veldist} we show the cumulative distribution of 
the X-ray sources relative to the cluster galaxies without 
X-ray emission. We plot separate velocity distributions for all cluster AGN 
more luminous than $L_X > 10^{41}$ \ergs\ and all more luminous than 
$L_X > 10^{42}$ \ergs. 

This figure demonstrates that the velocity distribution of cluster 
AGN is essentially identical to the other cluster members. A Kolmogorov-Smirnov 
(KS) test confirms this to be the case, specifically that there is a 68\% and 
96\% probability that the $L_X > 10^{41}$ \ergs\ and $L_X > 10^{42}$ \ergs\ 
AGN, respectively, are drawn from the same parent population 
as the cluster members without X-ray counterparts (see Table~\ref{tbl:ks}). 
We therefore do not find evidence that the cluster AGN are preferentially 
distributed on radial orbits, even if we exclude the merging clusters 
A3125/8, unlike the case for emission-line galaxies in clusters 
\citep{dressler99}. This may indicate that nuclear activity in cluster 
galaxies can remain, or be reactivated, after the parent population has 
virialized in the cluster potential. 

To test the sensitivity of our data to the known differences between 
cluster galaxy populations, we computed similar distributions between 
Butcher-Oemler and non-Butcher-Oemler galaxies as well as between 
emission-line and absorption-line galaxies. This analysis showed that the 
Butcher-Oemler galaxies have systematically larger velocity dispersions 
than the redder cluster members, as expected for a population on more 
radial orbits. The difference between emission-line and absorption-line 
cluster members was not statistically significant, although as noted 
previously the emission-line classification is more susceptible to variations 
in the average signal-to-noise ratio than the color classification. The 
emission and absorption-line samples used here are also only based on 
our spectroscopy, and not all known members, and therefore the sample is 
smaller than that used for the Butcher-Oemler and red galaxy comparison. 

  \subsection{Radial Distribution}  \label{sec:raddist} 

\begin{figure*}
\figurenum{7} 
\plotone{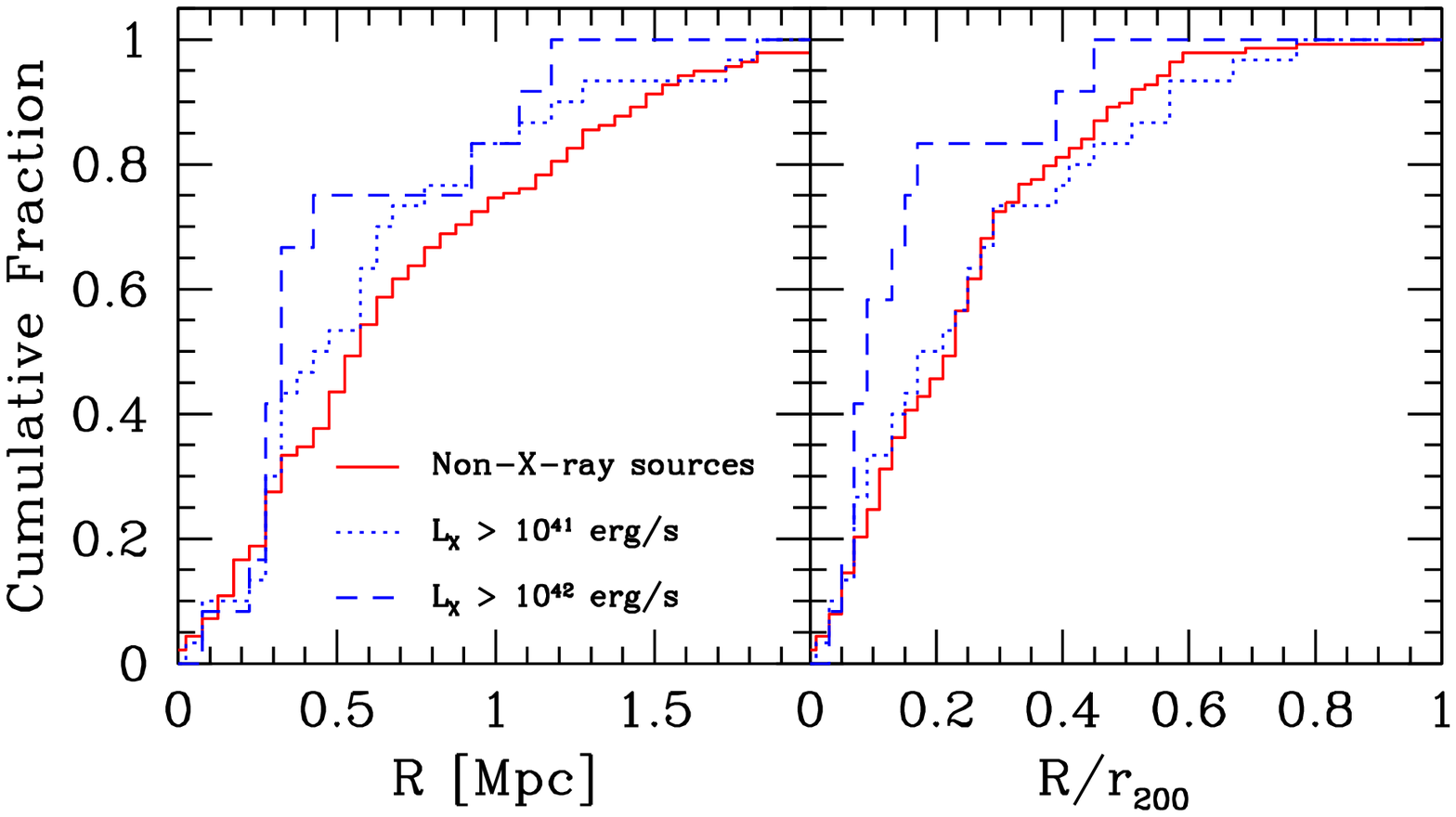} 
\caption{
Cumulative, normalized radial distributions relative to each cluster's center 
in projected Mpc ({\it left}) and in units of $r_{200}$ ({\it right}). 
The distribution of the AGN more luminous than $L_X > 10^{42}$ \ergs 
({\it blue, dashed line}) is substantially more centrally concentrated than 
the inactive galaxies ({\it red, solid line}) in both radial distributions. 
The distribution of the AGN more luminous than $L_X > 10^{41}$ \ergs 
({\it blue, dotted line}) is comparable to the inactive galaxies. 
The radial distribution is calculated from the projected distance of each 
galaxy from the cluster center. Larger separations are sampled by the more 
distant clusters. See \S\ref{sec:raddist} for further details. 
\label{fig:raddist} 
}
\end{figure*}

To investigate the radial distribution of the AGN relative to the cluster
center, we have similarly combined all of the cluster sources shown in
Figure~\ref{fig:pos} and plot the cumulative radial distribution of the AGN
and inactive cluster members in Figure~\ref{fig:raddist}. We have calculated
this distribution as a function of projected physical distance from the cluster
center in Mpc and as a function of projected fraction of $r_{200}$, where
$r_{200}$ is the physical radius within which the mean density of a virialized
cluster of the measured velocity dispersion exceeds the critical density at
that redshift by a factor of 200 \citep[\eg,][]{carlberg97,treu03}.
This figure shows that the $L_X>10^{41}$ \ergs\ cluster AGN and the other
cluster members have a similar radial distribution, but above
$L_X > 10^{42}$ \ergs\ {\it the cluster AGN are more strongly concentrated.}
A KS test confirms the visual impression of both panels of
Figure~\ref{fig:raddist}. There are formally only 7\% and 3\% probabilities
that the $L_X > 10^{42}$ \ergs\ AGN are drawn from the same distribution in
metric and virial radius, respectively (see Table~\ref{tbl:ks}).
The probabilities are less than a factor of 2 higher when we cluster the
merging clusters A3125/8.
This result confirms the previous, purely statistical evidence for a
concentration of AGN in the centers of clusters \citep{ruderman05,dowsett05}
with a spectroscopically-confirmed sample.
In a recent paper \citet{lin07} found evidence that radio sources in clusters
are also more centrally concentrated than typical cluster galaxies, however
we intriguingly do not find any evidence that the X-ray and radio AGN are in
the same host galaxies. Comparison of our X-ray selected AGN and those in the
radio study of \citet{morrison03} indicates that the two samples are nearly
disjoint (see also the radio and X-ray contours in Figure~\ref{fig:image1}).

The summed radial distribution of cluster members is more sensitive to
several potential biases than the summed velocity distribution. These
biases include the physical area of the cluster surveyed for X-ray and other
members and the relative physical sizes of the clusters. The first is
important because the distributions will artificially appear different if the
field of the X-ray observations is larger or smaller than the field of
view surveyed to obtain membership information for inactive galaxies.
We have addressed this source of systematic error by only including
our spectroscopically confirmed members in Figure~\ref{fig:raddist},
rather than all available members. Our spectroscopic sample is
a suitable comparison sample because the X-ray and other cluster
galaxy candidates were observed with the same multislit masks and
the AGN and inactive galaxies in each cluster were targeted over the
same radial distribution from the cluster center. We further refined the
input catalog by only including confirmed cluster members that fall
within the field of view of the \chandra\ observations.

The second potential bias is the different sizes of the clusters, both
their physical size and projected angular extent relative to the \chandra\
field of view. For example, the ACIS-I camera on \chandra\ only encompasses
the central $\sim 0.5$ Mpc of Abell~3125 and Abell~3128, while it encompasses
up to $\sim 3$ Mpc from the center of MS1008 (see Figure~\ref{fig:pos}).
The redshift range of the sample therefore causes the higher-redshift clusters
to dominate the distribution at the largest distances from the cluster center.
Similarly, only the low-mass cluster A3125 and the higher-redshift
cluster MS1008 provide substantial information about the distribution of
cluster galaxies at more than half of $r_{200}$. The virial
scaling does not change the fact that the \chandra\ field of view samples a
a different fraction of the physical extent of different clusters.

A final potential bias on the radial distribution of X-ray sources is the
detection efficiency of X-ray sources as a function of distance from the
cluster center. The detection efficiency may be a function of clustercentric
distance due to two effects: the higher background near the cluster center due
to the ICM (see Figures~\ref{fig:image1} and \ref{fig:image2}) and the
degradation of the \chandra\ point-spread function (PSF) at larger off-axis 
angles may make it more
difficult to detect sources. While the effect of the bright ICM would
decrease the probability of detecting an X-ray source near the cluster
core and therefore is opposite our observed trend, the larger \chandra\
PSF at larger off-axis angles could mimic the observed trend for
$L_X > 10^{42}$ \ergs\ AGN.

To quantify both of these potential biases on the apparent radial distribution,
we performed a series of simulations with the
MARX\footnote{http://space.mit.edu/CXC/MARX} package to quantify our detection
efficiency as a function of source luminosity and clustercentric distance.
MARX is well-suited to this task because it can generate a PSF that accounts
for the off-axis angle, aspect solution, and spectral shape. The detection
efficiency relative to the observed radial distribution of all members with
$M_R < -20$ mag and no X-ray detection is the most relevant quantity because
our comparison is directly with these galaxies. We therefore used MARX to add
an artificial X-ray point source at the location of every cluster member
without an X-ray source. This source was assigned a $\Gamma = 1.7$ spectra
shape and the total counts corresponding to a given luminosity at the
redshift of the particular cluster.
The fraction of these sources recovered with the same {\sc wavdetect}
parameters used in \citet{martini06} then provides a measurement of the
detection efficiency. We performed this exercise for point sources with
total counts corresponding to a source with $L_X = 10^{41}$ and
$L_X = 10^{42}$ \ergs\ for each cluster and then calculated the detection
efficiency as a function of clustercentric distance in both Mpc and
fraction of the virial radius. From this approach we find that the detection
efficiency varies by less than 20\% out to 1.5Mpc and 0.5$r_{200}$, after which
the results begin to be compromised by small number statistics (few cluster
members). This is true for both $10^{41}$ and $10^{42}$ \ergs\ sources and
we therefore conclude that the observed trend with radius is real and not an
artifact of either bias. We note further that this approach is somewhat
pessimistic as it assumes that all of the point sources are at the luminosity
limit, rather that at or above it, but it nevertheless provides a relative
measure of the detection probability.

As in \S~\ref{sec:veldist}, we compared the radial distributions of
photometrically (Butcher-Oemler or red) and spectroscopically
(emission-line or absorption-line) defined subsamples to test the
sensitivity of these data to other known relations between cluster
populations. This analysis showed that the red galaxies are more
centrally-concentrated than the Butcher-Oemler population, as expected.
We also found no statistically-significant difference between the
emission-line and absorption-line populations, which we again
attribute to the smaller size of the subsamples input to this analysis.

  \subsection{Association with Cluster Substructure} 

\begin{figure*}
\figurenum{8} 
\plotone{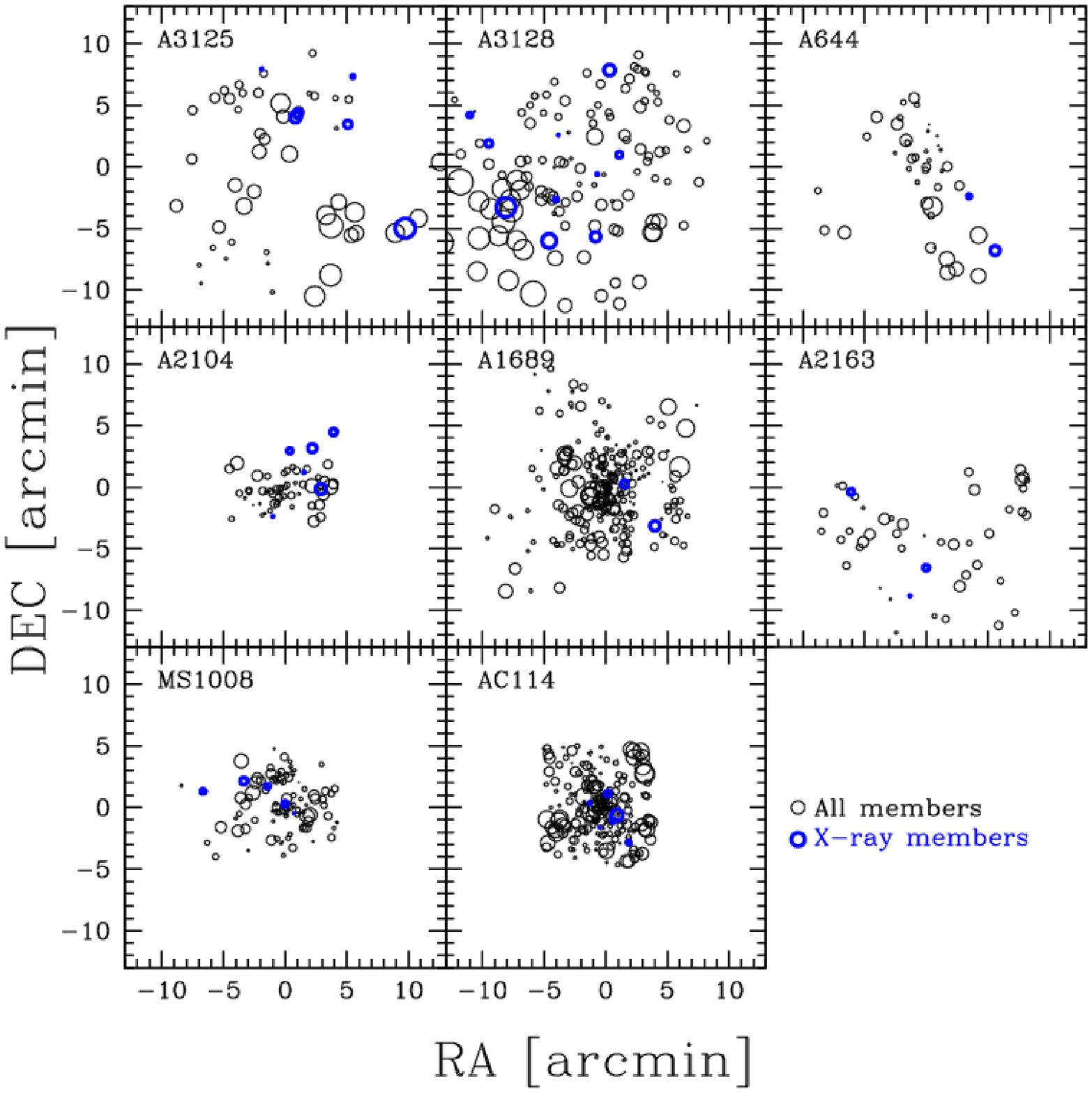} 
\caption{
\label{fig:sub} 
Cluster substructure calculated from known cluster members. 
Positions of all cluster members ({\it open circles}) and X-ray members 
({\it thick, blue circles}) are relative to the centers listed in 
Table~\ref{tbl:clusters}. The size of each circle is scaled by the value 
of the substructure parameter $\delta$ for each source. 
The axes on all panels are offsets in arcminutes from the cluster center. 
}
\end{figure*}

If cluster AGN are preferentially associated with infalling groups of galaxies, 
or other structures within clusters, then they may trace local deviations from 
the mean cluster kinematics. \citet{dressler88} developed a substructure test 
to identify deviations in the cluster velocity field, where the substructure 
parameter $\delta$ is defined as 
\begin{equation} 
\delta^2 = (11/\sigma^2) \left[ (\bar{\upsilon}_{local} - \bar{\upsilon})^2 + (\sigma_{local} - \sigma)^2 \right]
\end{equation} 
and the local values of the radial velocity $\bar{\upsilon}_{local}$ and 
velocity dispersion $\sigma_{local}$ are calculated from the radial velocities 
of each galaxy and its 10 nearest neighbors. Larger values of $\delta$ 
correspond to larger local deviations from the cluster kinematics. 
Figure~\ref{fig:sub} plots all of the cluster 
members shown in Figure~\ref{fig:pos} with open circles scaled by this 
substructure measure. 
One minor change from the method of \citet{dressler88} is that we calculate 
these quantities with the biweight estimator described previously, rather than 
assume a Gaussian distribution. 

This analysis indicates that Abell~3125 and Abell~3128 have the most 
substantial substructure. \citet{dressler88} note that the sum of the $\delta$ 
values for all cluster members is a measure of how much substructure is present 
and a simple way to characterize this value for a given cluster is via Monte 
Carlo simulations with random reshuffling of the measured velocities among the 
measured positions. We have generated 10,000 realizations with random 
redistributions of the measured velocities among the measured galaxy positions 
and calculated the probability of obtaining the measured sum of the $\delta$ 
values or larger. These values are listed in Table~\ref{tbl:clusters} and 
indicate that Abell~3125 and Abell~3128 indeed have significant substructure, 
which is not surprising for a merging pair of clusters, while the remaining 
clusters do not have substantial substructure. 

\begin{figure}
\figurenum{9} 
\plotone{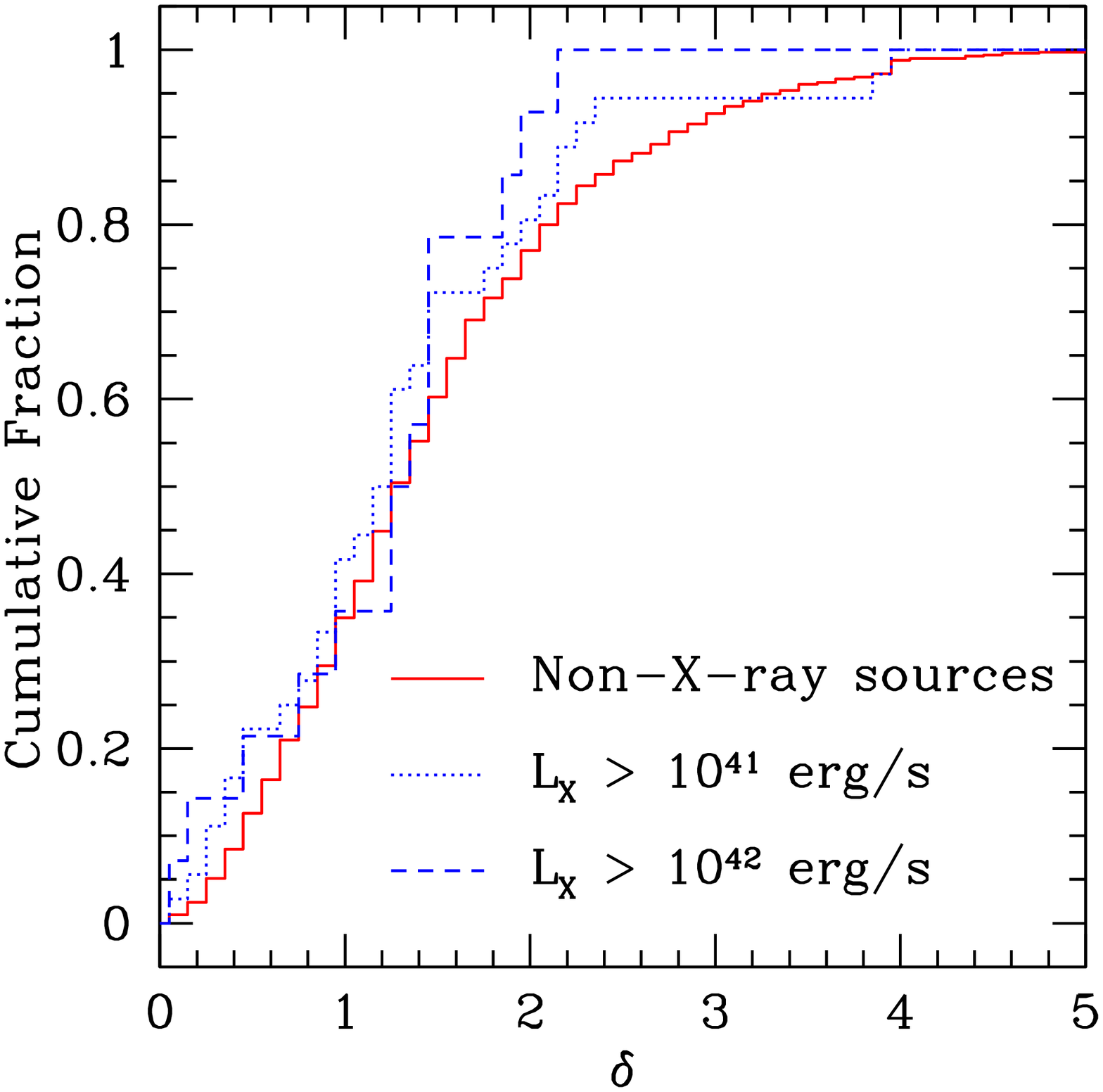} 
\caption{
\label{fig:delta} 
Cumulative distribution of the substructure parameter $\delta$ for X-ray 
sources more luminous than $L_X > 10^{41}$ \ergs ({\it blue, dotted line}), 
more luminous than $L_X > 10^{42}$ \ergs ({\it blue, dashed line}), and 
inactive cluster members ({\it red, solid line}). Larger $\delta$ corresponds 
to a larger local deviation from the mean cluster kinematics. 
}
\end{figure}

The substructure parameter can also be used to characterize the extent that 
a particular cluster population is associated with substructure. We have 
calculated the distribution of $\delta$ values for the X-ray sources in 
all eight clusters relative to the other cluster members and plot this 
distribution in Figure~\ref{fig:delta}. There is good agreement between 
the distribution of the $L_X>10^{41}$ \ergs, $L_X>10^{42}$ \ergs, and 
cluster members without X-ray counterparts, and this indicates that the 
cluster AGN are not more likely to be associated with substructure in 
clusters, at least on the scales this parameter is sensitive to (a galaxy and 
its ten nearest neighbors). 
As Abell~3125 and Abell~3128 have substantial substructure, we 
repeated this exercise without these two cluster and still obtained consistent 
distributions. Results from KS tests of all of these distributions are 
listed in Table~\ref{tbl:ks}. 

\section{Relation of AGN to Cluster Properties} \label{sec:relation} 

In the previous two sections we have analyzed the properties of the 
clusters and the distribution of the AGN within the clusters. In the 
present section we discuss the completeness of our survey of each cluster 
and investigate potential correlations between the AGN fraction and the 
properties of their host cluster. 

  \subsection{Completeness}  \label{sec:complete}

\begin{deluxetable*}{lccccccccc}
\tablecolumns{10}
\tablewidth{7.0truein}
\tabletypesize{\scriptsize} 
\tablecaption{Completeness and AGN Fractions \label{tbl:fa}}
\tablehead{
\colhead{Cluster} &
\colhead{$N_X$} &
\multicolumn{3}{c}{-------- $N_X$ ($<-20$ mag) --------} &
\multicolumn{2}{c}{------ Membership ------} &
\multicolumn{1}{c}{} &
\multicolumn{2}{c}{-------- $f_A$ --------} \\
\colhead{Name} &
\colhead{} &
\colhead{All} &
\colhead{$>10^{41}$} &
\colhead{$>10^{42}$} &
\colhead{Confirmed} &
\colhead{Estimated} &
\colhead{All} &
\colhead{$>10^{41}$} &
\colhead{$>10^{42}$} \\
\colhead{(1)} &
\colhead{(2)} &
\colhead{(3)} &
\colhead{(4)} &
\colhead{(5)} &
\colhead{(6)} &
\colhead{(7)} &
\colhead{(8)} &
\colhead{(9)} &
\colhead{(10)} \\
}
\startdata
 Abell3125 &   6 &   5 &   4 &   0 &   18 &   28 & 0.179 & 0.143 & 0.000 \\ 
 Abell3128 &  10 &  10 &   7 &   1 &   54 &   63 & 0.159 & 0.111 & 0.016 \\ 
  Abell644 &   2 &   2 &   2 &   1 &   15 &   75 & 0.027 & 0.027 & 0.013 \\ 
 Abell2104 &   6 &   4 &   4 &   1 &   33 &   62 & 0.065 & 0.065 & 0.016 \\ 
 Abell1689 &   2 &   2 &   2 &   2 &  110 &  337 & 0.006 & 0.006 & 0.006 \\ 
 Abell2163 &   3 &   2 &   2 &   1 &   25 &  262 & 0.008 & 0.008 & 0.004 \\ 
    MS1008 &   5 &   4 &   4 &   3 &   75 &  346 & 0.012 & 0.012 & 0.009 \\ 
     AC114 &   6 &   5 &   5 &   3 &   95 &  204 & 0.025 & 0.025 & 0.015 \\ 
           &     &     &     &     &      &      &       &       &       \\ 
   Average &     &     &     &     &      &      & 0.060 & 0.049 & 0.010 \\ 
       Sum &  40 &  35 &  30 &  12 &  425 & 1377 & 0.025 & 0.022 & 0.009 \\ 
\enddata
\tablecomments{
Completeness estimates and AGN fractions for galaxies more luminous than 
$M_R = -20$ mag and different X-ray luminosity cuts. Columns are: 
(1) Cluster name; (2) Number of X-ray sources in the cluster regardless of 
the host galaxy luminosity; (3) Number of X-ray sources in hosts more luminous 
than $M_R = -20$ mag; (4) Number of X-ray sources more luminous than $L_X = 
10^{41}$ \ergs\ in the broad X-ray band; (5) Number of X-ray sources more 
luminous than $L_X = 10^{42}$ \ergs; (6) Number of spectroscopically confirmed 
cluster members within the \chandra\ field of view more luminous than $M_R = 
-20$ mag; (7) Estimate of the total number of members more luminous than 
$M_R = -20$ mag; (8) AGN fraction if all X-ray sources are AGN; (9) AGN 
fraction if all X-ray sources more luminous than $L_X = 10^{41}$ \ergs\ are 
AGN; (10) same as column (9) for X-ray sources more luminous than $L_X = 
10^{42}$ \ergs.  The final two rows present the AGN fraction calculated from 
the average of the eight clusters and the average of the sum of the eight 
clusters.  See \S~\ref{sec:complete} for further details. 
}
\end{deluxetable*}

In Paper~I we calculated the AGN fraction in clusters of galaxies and 
defined the AGN fraction as the fraction of galaxies more luminous than a 
rest-frame $M_R = -20$ mag with X-ray counterparts more luminous than 
$L_X = 10^{41}$ \ergs\ in the broad X-ray band, or $f_A(M_R<-20;L_X>10^{41}) 
= 5 \pm 1.5$\%. The quoted uncertainty in this value was only a Poisson 
estimate and did not take into account potential systematic uncertainties, 
such as the estimate of the total cluster galaxy population. 
The determination of the numerator in the AGN fraction 
requires redshift measurements for all X-ray counterparts brighter than 
the apparent magnitude of a cluster member with this absolute magnitude. As we 
discussed in \S\ref{sec:obs}, our spectra are complete to the 
appropriate apparent magnitude limit for each cluster. Measurement of the 
denominator requires a much larger number of spectra because most galaxies to 
the requisite apparent magnitude limit in the \chandra\ field of view are not 
detected X-ray sources. For all but the lowest-redshift clusters, these data 
are substantially incomplete and we have to sum the known members with an 
estimate of the additional cluster members without redshifts to estimate the 
total number of cluster members brighter than $M_R = -20$ mag. As noted in 
Paper~I, we can quantify this completeness with photometric observations of 
known cluster members, known nonmembers, and galaxies without spectra. 
Here we estimate the total number of cluster members without spectra by 
calculating the fraction of galaxies with spectra that are cluster members 
as a function of $B-R$ color, rather than adopt a color-independent 
completeness estimate as in Paper~I. This color-dependent completeness estimate 
takes into account the observation that galaxies with similar colors to 
observed cluster members are more likely to be cluster members than those with 
dramatically different colors, but still accounts for a potential population 
of bluer or Butcher-Oemler galaxies if such a population has been 
spectroscopically confirmed. A disadvantage of this approach is that the 
completeness estimate is sensitive to the number of members and nonmembers 
with spectra at a given color. We have increased the number of galaxies with 
spectra at a given color by also including known members and nonmembers from 
the literature for which we have photometry, although they fall outside of the 
\chandra\ field of view. However, these additional data are only used to 
provide an improved completeness fraction as a function of color; the total 
number of cluster members is estimated only from galaxies that fall within the 
\chandra\ field of view. 

In Table~\ref{tbl:fa} we use the results of this completeness calculation
to obtain an estimate of the total number of cluster members more luminous than 
$M_R = -20$ mag. For each cluster we list both the number of known members 
from our data and the literature that fall within the \chandra\ field of view 
and our estimate of the total cluster galaxy population. We then calculate 
the AGN fraction for these clusters with three different cuts in X-ray 
luminosity: (1) the fraction of AGN if all of the X-ray sources in galaxies 
with $M_R < -20$ mag identified in Paper~I are AGN; (2) the AGN fraction for 
only those X-ray 
sources above a luminosity of $L_X = 10^{41}$ \ergs; and (3) the AGN fraction for 
those above $L_X = 10^{42}$ \ergs. The second of these three limits, namely 
$M_R < -20$ mag and $L_X > 10^{41}$ \ergs\ corresponds to the AGN fraction 
quoted in Paper~I. Here we derive $f_A(M_R<-20;L_X>10^{41}) = 5$\%, 
in excellent agreement with our calculation from Paper~I. 
This correction provides a better estimate of the membership fraction for 
galaxies observed to be redder and bluer than typical, known members. The X-ray 
luminosity threshold was chosen previously because it is rare for other 
potential sources of X-ray emission, most notably LMXBs and hot, gaseous 
halos, to produce such high X-ray luminosities. However, as noted in Paper~I, 
the disadvantage of this threshold is that the X-ray observations of 
Abell~1689 and MS1008 are not quite sensitive to this luminosity limit and 
therefore some AGN may be missed and the AGN fraction 
underestimated. We have also calculated the AGN fraction for a factor of ten 
higher X-ray luminosity and find $f_A(M_R<-20;L_X>10^{42}) = 1$\%.  All of 
our X-ray observations are sensitive to such luminous sources. Galaxies 
that can produce such luminous X-ray emission by mechanisms other than black 
hole accretion are also much rarer than AGN. 

The AGN fraction may also be a function of host galaxy luminosity, in addition 
to X-ray luminosity. In our original work on Abell~2104 \citep{martini02}, we 
noted that two of the three brightest cluster members were AGN. From 
inspection of Figure~\ref{fig:cmd}, a similar tendency is apparent at the 
bright end of the color-magnitude relation in Abell~3125, Abell~3128, and 
Abell~2163. In fact, approximately half (19) of the AGN more luminous than 
$L_X > 10^{41}$ \ergs\ are in host galaxies more luminous than $M_R<-21.3$ mag 
and for these galaxies we find $f_A(M_R<-21.3;L_X>10^{41}) = 9.8$\%. 
This absolute 
magnitude limit was chosen to correspond to the luminosity threshold of 
\citet{sun07}, who find nine AGN with $L_X > 10^{41}$ \ergs\ in 163 galaxies 
in a sample of $z = 0.01 - 0.05$ clusters, or 
$f_A(M_R<-21.3;L_X>10^{41}) = 5$\%. While this value is a factor of two 
below our value, the differences may be due to variations in the AGN 
fraction between cluster samples (e.g. different redshift distributions) and 
we discuss this point further in \S\ref{sec:var}. 

An important caveat to these estimates is that they are estimates of the 
AGN fraction within the field of view of the \chandra\ ACIS-I or ACIS-S 
detectors. These detectors have an approximate field of view of $16.9' \times 
16.9'$ and $8.3' \times 8.3'$, respectively. These two angular sizes, combined 
with the redshift distribution of our sample and the range in cluster sizes 
(\eg, see the $r_{200}$ value in Table~\ref{tbl:clusters}), 
correspond to substantial variations in the fraction of each cluster galaxy 
population surveyed. This point is illustrated with Figure~\ref{fig:pos},
which includes circles that mark the $r_{200}$ radius. The \chandra\ field of 
view approximately covers out to the $r_{200}$ radius for only MS1008, while 
for many of the remaining clusters it is even larger than the $25'$ square 
panels shown for each cluster. Our measurements of the AGN fraction for 
each cluster are therefore necessarily measurements dominated by the 
center of the cluster for nearly all cases. However, the absence of a strong 
radial trend in the AGN distribution for the more common $L_X > 10^{41}$ 
\ergs\ sources shown in Figure~\ref{fig:raddist} 
suggests that the ratio of AGN to inactive galaxies may not be a strong 
function of position within the cluster. The AGN fraction may therefore be 
relatively well determined, even if the census of the total cluster AGN 
population is incomplete.  We also note that for several 
clusters our ground-based images do not encompass the entire \chandra\ 
field, and in fact some of the X-ray sources identified in A3125 and A3128 
are from literature positions and redshifts, but the measurement of the 
AGN fraction remains relatively robust in the absence of strong 
radial gradients. 

We have assumed that each cluster of galaxies has the same, intrinsic AGN 
fraction (and the same X-ray luminosity function) and treated each cluster 
as an independent measure of the AGN fraction. This yields an average fraction 
of $5$\% for $L_X > 10^{41}$ \ergs\ and $1$\% for $L_X > 10^{42}$ \ergs. 
If every cluster galaxy had an equal probability of hosting an AGN, an 
alternate way to calculate the AGN fraction in clusters of galaxies would be 
to treat each cluster galaxy as an independent measure of the AGN fraction, 
rather than each cluster. This yields an AGN fraction of $2.2$\% for 
$L_X > 10^{41}$ \ergs\ and $0.9$\% for  $L_X > 10^{42}$ \ergs. The value for 
$L_X > 10^{41}$ \ergs\ sources is substantially lower than the average of the 
cluster AGN fractions and provides evidence that the AGN fraction may vary 
from cluster to cluster. 

  \subsection{Evidence for Intrinsic Variation} \label{sec:var} 

We now investigate variations in the AGN fraction from cluster to cluster and 
consider if these variations are consistent with Poisson fluctuations 
and systematic errors in the cluster membership. 
If we ignore potential systematic errors in the completeness correction, 
the AGN fractions in these eight clusters are inconsistent with Poisson 
fluctuations with greater than 95\% confidence for $L_X > 10^{41}$ \ergs; 
there is no evidence for variations in the AGN fraction for the much 
smaller population with 
$L_X > 10^{42}$ \ergs. The dominant source of systematic error in the AGN 
fraction is the estimate for the number of cluster galaxies in the \chandra\ 
field of view. This correction is more important for clusters that have a 
larger number of galaxies within the field of view of the \chandra\ 
observation and those clusters that have less membership data. The ratio of 
the estimated to confirmed cluster galaxy populations brighter than 
$M_R < -20$ mag in Table~\ref{tbl:fa} provides the size of the correction and 
varies from $\sim 1$ for Abell~3128 to $\sim 10$ for A2163 with a median of 
$\sim 2.5$. The ratio of the number of AGN to the number of confirmed 
members provides a strong upper limit to the AGN fraction in each cluster.  
If we use this larger value for the clusters with the smallest AGN fraction, 
rather than the value listed in Table~\ref{tbl:fa}, we still find the 
AGN fraction in these clusters is inconsistent with the values derived for 
the lower-redshift clusters with small systematic errors. 
For example, the well-studied, high-redshift cluster AC114 has an AGN fraction 
of $5.2$\% 
based on confirmed members alone and this upper limit is still inconsistent 
with the AGN fraction in clusters such as Abell~3125 and Abell~3128 that have 
small completeness corrections. 
The variation in the AGN fraction from cluster to cluster also does not seem to 
be due to radial dependence in the AGN fraction, as Figure~\ref{fig:raddist} 
indicates that there is not a difference between the distributions of the 
$L_X > 10^{41}$ \ergs\ population and other cluster members. 

  \subsection{Correlations with Cluster Properties} 

\begin{figure*}
\figurenum{10} 
\plotone{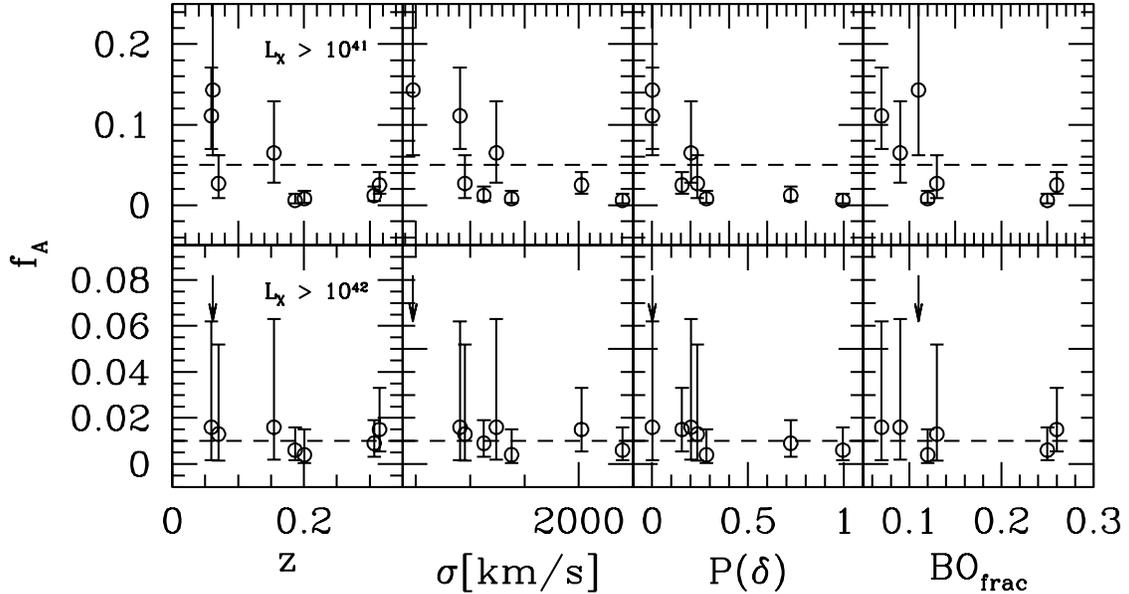} 
\caption{
AGN fraction as a function of redshift, cluster velocity dispersion, amount 
of cluster substructure, and estimated fraction of Butcher-Oemler galaxies. 
The AGN fraction for each cluster is listed in columns 9 and 10 of 
Table~\ref{tbl:fa} and the error bars are Poisson 90\% confidence limits. 
The top row contains AGN with $L_X > 10^{41}$ erg s$^{-1}$ and the bottom row 
contains AGN with $L_X > 10^{42}$ erg s$^{-1}$. 
The horizontal, dashed line corresponds to the average AGN fraction of the 
eight clusters for each luminosity threshold. 
The cluster properties are provided in Table~\ref{tbl:clusters}. 
\label{fig:fa} 
}
\end{figure*}

As there is evidence for variation in the AGN fraction from cluster to 
cluster, we compare the AGN fraction from each cluster with four properties 
that may influence the AGN fraction: cluster redshift; cluster velocity 
dispersion; amount of cluster substructure; and the fraction of Butcher-Oemler 
galaxies in the cluster. Simple arguments about the nature of AGN fueling 
and known relations between AGN and their host galaxies in the field motivates 
our choice of these properties. First, a dependence with redshift is expected 
because of the pronounced decrease in the space density of AGN from 
high redshift to the present \citep[e.g.,][]{osmer04}. In addition, the 
massive, early-type galaxies in clusters should harbor supermassive black 
holes that most likely grew by accretion as luminous AGN at earlier times. 
Statistical studies of X-ray point source overdensities toward clusters do 
show evidence of an increase in the AGN fraction in high-redshift clusters 
\citep{dowsett05}, although these AGN have not been spectroscopically 
confirmed to be cluster members. 
Nevertheless, these arguments suggest that the AGN population in clusters of 
galaxies should be higher at higher redshift, although they do not necessarily 
require pronounced evolution over the redshift range of this sample. 
In particular, most of the AGN in these clusters are low luminosity 
($L_X \sim 10^{41-42}$ \ergs) and current cosmic downsizing models indicate 
that there is relatively little evolution at these luminosities at low redshift 
\citep[e.g.,][]{ueda03,hasinger05}. 
In fact the upper-left panel of Figure~\ref{fig:fa} actually shows that the 
AGN fraction increases at {\it lower} redshift for the $L_X>10^{41}$ \ergs\  
sample, contrary to these expectations. 
However, there are substantial correlations between redshift and several 
other properties of these clusters and therefore this apparent correlation 
may be a selection effect. 
The most important of the potential systematic effects produced by these 
correlations are due to the two lowest-redshift clusters, the merging pair 
Abell~3125/3128. These two clusters possess the greatest substructure and 
also have the lowest velocity dispersions. The cluster sample for this survey 
was simply chosen from available \chandra\ data in the southern hemisphere to 
obtain a measurement of the AGN fraction in clusters \citep[see ][]{martini06} 
and these potential parameter correlations were not foreseen. 

Perhaps the most important correlation between two cluster properties in our 
sample is that between redshift and velocity dispersion. If mergers play an 
important role in triggering low-luminosity AGN, we expect that clusters 
of galaxies with lower velocity dispersions will have a higher AGN fraction 
because the merger rate will increase. The second column of Figure~\ref{fig:fa} 
shows that some of the lower velocity dispersion clusters do have higher AGN 
fractions, although this trend is largely driven by Abell~3125/3128. 
Recent evidence for an anticorrelation between AGN fraction and velocity 
dispersion comes from the work of \citet{popesso06}, who studied the AGN 
fraction as a function of velocity dispersion with SDSS spectroscopic data. 
These authors found that the average AGN fraction increased in environments 
with lower velocity dispersion. 

Finally, the last two columns of Figure~\ref{fig:fa} show the AGN fraction 
as a function of cluster substructure P($\delta$) and the fraction of 
Butcher-Oemler galaxies in the cluster. The AGN fraction may depend on 
cluster substructure if cluster-cluster mergers produce a substantial 
increase in the galaxy interaction rate. In the above discussion of how the 
merger rate is lower in higher velocity-dispersion environments, we neglected 
the potential role of fast fly-by interactions on fueling AGN. 
These fast interactions have been studied as a mechanism for driving galaxy 
evolution in clusters \citep{moore96,gnedin03} and may also provide a 
mechanism for fueling AGN \citep{lake98}. While it is not clear if the 
rate of fly-by interactions will increase in clusters with substantial 
substructure, our measurements do indicate that the two merging clusters 
have the highest AGN fraction. Surprisingly, these clusters also have 
among the lowest fraction of Butcher-Oemler galaxies and this produces an 
apparent trend that clusters with fewer starforming galaxies have a larger 
AGN fraction. While again this is based on a small sample that consequently 
does not merit too much interpretation, this is contrary to our expectation 
that clusters with more star formation and more available cold gas would be 
those with higher AGN fractions. 

\section{Summary} \label{sec:conclude} 

We have investigated the distribution of AGN in clusters of galaxies in order 
to study AGN and galaxy evolution in rich environments. Specific questions 
that motivated our study include the following: What mechanisms fuel accretion  
onto supermassive black holes? Is the evolution of AGN in clusters different 
from AGN evolution in the field? We have taken a step toward answering 
these questions with a study of the relative distributions of AGN and 
inactive galaxies in clusters and how the cluster AGN fraction varies as a 
function of several cluster properties. Our main results are as follows: 

1. The most luminous AGN ($L_X > 10^{42}$ \ergs) are more centrally 
concentrated than cluster galaxies with similar absolute magnitude 
$M_R < -20$ mag, while less luminous AGN ($L_X > 10^{42}$ \ergs) have a 
similar distribution to other cluster members. This greater central 
concentration of the most luminous members is contrary to our expectation 
that most AGN will be triggered in the outskirts of clusters. While this 
survey does not extend to large enough radius to sample the cluster outskirts, 
our data do demonstrate that the $L_X > 10^{42}$ \ergs\ AGN are more centrally 
concentrated in the cluster than inactive galaxies of the same absolute 
magnitude and these AGN are not dominated by galaxies that have not recently 
fallen into the cluster. We would see newly infalling galaxies in projection 
in the kinematic distribution; however, as the cluster AGN have a similar 
kinematic distribution to other cluster galaxies, this suggests that they are 
not significantly more or less likely to be on a radial orbit than a typical 
cluster galaxy. Cluster AGN and inactive cluster galaxies also similarly 
trace cluster substructure. Future observations that encompass a larger 
fraction of the virial radial will help to physically separate AGN in the 
cores of clusters from any potential AGN population in the cluster outskirts. 

2. The AGN fraction in clusters varies between clusters to a greater 
extent than can be explained by Poisson statistics alone. Specifically, 
the estimated AGN fraction $f_A(M_R<-20;L_X>10^{41})$ in cluster galaxies 
ranges from 14\% to 0.6\% with a mean fraction of 5\% per cluster. 
While there are some systematic uncertainties due to our estimates of the 
cluster completeness, the statistical and systematic uncertainties for 
several clusters remain inconsistent with one another with greater than 
90\% confidence. We examined if the variation in AGN fraction from cluster to 
cluster correlates with any cluster property and found weak evidence that the 
AGN fraction is higher at lower redshift, in lower velocity-dispersion 
clusters, in clusters with substantial substructure, and in clusters with a 
smaller fraction of Butcher-Oemler galaxies. Unfortunately, our sample of 
clusters is relatively small and has some substantial correlations between 
these properties. In particular, the two lowest-redshift clusters are also 
those with the most substructure and the lowest velocity dispersions 
(Abell~3125 and Abell~3128). Therefore we cannot conclude which of 
these properties, or what combination of these properties, sets the AGN 
fraction in clusters. A larger sample of clusters that more evenly 
fills this parameter space is needed to determine what parameters set the 
AGN fraction in clusters of galaxies. These data could then be used to 
test models of AGN fueling, for example via the functional form of the 
scaling between AGN fraction and velocity dispersion. 

Together these two results provide evidence that the properties of cluster AGN, 
like the properties of cluster galaxies, vary both within clusters 
and from cluster to cluster. An equally interesting question to answer is 
if the AGN fraction varies substantially between clusters and lower-density 
groups and the field. While evidence from spectroscopic surveys unequivocally 
answers yes to this question and demonstrates that AGN are less commonly found 
in clusters than the field, radio and now our X-ray observations indicate that 
the dependence of AGN fraction on environment may be much less stark when the 
AGN are selected at these other wavelengths. While differences in the typical 
spectral properties of AGN as a function of environment does not affect the 
present comparison between and within clusters, they will complicate 
comparisons of the AGN fraction between the field, groups, and clusters of 
galaxies.  A possible resolution is that the AGN fraction may be a function 
of morphological type and not just absolute magnitude. This is 
known to be the case for spectroscopically-classified AGN \citep{ho97e} in 
the sense that early-type galaxies have a higher incidence of AGN. 
As \citet{lehmer07} measured a comparable X-ray selected AGN fraction in 
morphologically-classified early-type galaxies, and these dominate in clusters 
but not groups, future measurements of the AGN fraction may need to consider 
morphological type, in addition to host luminosity and AGN luminosity. 
This may reflect the fact that morphological type sets the spheroid mass and 
consequently the black hole mass for a galaxy of a given total luminosity. 

\acknowledgements 

We would like to thank R. Gilmour (n\'ee Dowsett) for helpful discussions
and acknowledge B. Lehmer and M. Sun for providing us with more information 
about their measurements of the AGN fraction. 
Support for this work was provided by the National Aeronautics and Space 
Administration through Chandra Award Numbers 04700793 and 05700786 issued by 
the Chandra X-ray Observatory Center, which is operated by the Smithsonian 
Astrophysical Observatory for and on behalf of the National Aeronautics 
Space Administration under contract NAS8-03060.
We greatly appreciate the excellent staffs of the Las Campanas Observatory 
and the Magellan Project Telescopes for their assistance with these 
observations.
This paper includes data gathered with the 6.5 meter Magellan Telescopes 
located at Las Campanas Observatory, Chile.
This research has made use of the NASA/IPAC Extragalactic Database (NED) 
which is operated by the Jet Propulsion Laboratory, California Institute of 
Technology, under contract with the National Aeronautics and Space 
Administration.

{\it Facility:} {\facility{du Pont (Tek No. 5 imaging CCD, WFCCD)}, 
\facility{Magellan:Clay (LDSS2 imaging spectrograph)}

\end{document}